\documentclass[amssymb,floatfix]{revtex4}
\usepackage{graphicx}
\usepackage{dcolumn}
\usepackage{bm}
\begin{document}
\title{Tuning of bound states in the continuum by waveguide rotation}
\author{Almas F. Sadreev,  Artem S. Pilipchuk, and Alina A. Lyapina}
\address{Kirensky Institute of Physics, Federal Research
Center KSC SB RAS, 660036 Krasnoyarsk, Russia}
\date{\today}
\begin{abstract}

We consider acoustic wave transmission in non axisymmetric
waveguide which consists of cylindrical resonator and two
semi-infinite cylindrical waveguides whose axes are shifted
relative to the resonator axis  and each other by azimuthal angle
$\Delta\phi$. We show that for rotation of one of attached
waveguides the coupling matrix elements of the eigenmodes of
resonator classified by the integer $m$ and propagating
mode of the waveguide acquire phase factor $e^{im\Delta\phi}$.
That crucially effect Fano resonances  and creates an analog of
faucet opening and closing wave flux under rotation of the
waveguide. We show that under the rotation of the waveguide and
variation of the length of resonator numerous bound states in the
continuum occur complimenting by the Fano resonance collapse.
\end{abstract}
 \maketitle
\maketitle
\section{Introduction}
\begin{figure}
\includegraphics[width=12cm,clip=]{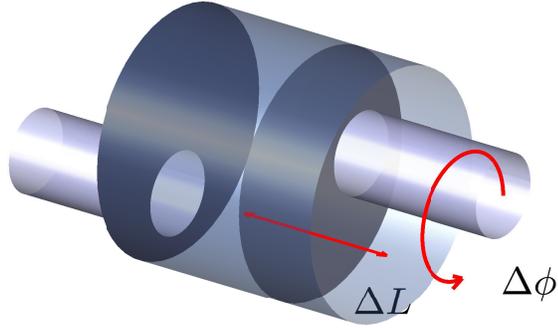}
\caption{(Color online)  Cylindrical resonator of radius $R=3$ and
length $L$ with two non-axially
 attached cylindrical waveguides of radius $r=1$. The input waveguide can freely
move along and rotate.}
\label{fig1}
\end{figure}

The beauty of complex wave fields observed in different areas of
physics is related to the phase of the field. The first
demonstration of the phase effect was done by Thomas Young in 1803
in the historic experiment on double-slit
interference\cite{Young}. Among numerous phase features we mark
bound states in the continuum (BSC) \cite{Neumann}, Aharonov-Bohm
oscillations \cite{ABO}, Fano asymmetric resonance \cite{Fano},
and topological singularities \cite{Nye} which milestone the phase
features in the last century.
In the present paper we consider a waveguide setup which unites
these features in wave transmission. The setup as shown in Fig.
\ref{fig1} consists of a cylindrical resonator of radius $R$ and
length $L$ with two attached semi-infinite cylindrical waveguides.
Tuning of the shape of the resonator could be performed in a
realistic acoustic or electromagnetic experiment by the use of
piston-like hollow-stem waveguides tightly fit to the interior
boundaries of a cylindric cavity \cite{Lyapina} as shown in Fig. \ref{fig1}.

If the waveguides were attached coaxially the total system would
be invariant relative to azimuthal rotation to split the total
Hilbert space of the system into independent subspaces given by
the orbital angular momentum (OAM) $m$. Respectively the wave
transmission would take place independently in each sector given
by the integer $m$ with prohibition for OAM conversion. Variation
of the resonator length rearranges the eigenlevels  to allow for
tuning Fano resonances which can be distinguished by changing the
coupling strengths of the eigenmodes with the propagating modes of
waveguide \cite{Kuhl}.

In the present paper we choose different strategy for tuning of Fano resonances
based on an attachment of the phase to the coupling strengths by means of the rotation
of one of the waveguides by the angle $\Delta\phi$ as shown in Fig. \ref{fig1}.
Then one of the waveguides acquires
phase difference relative to the other that crucially effects interference of
resonances, i.e., the Fano resonances and the wave transmission. We show that even tiny
rotations result in change of the transmittance from zero to the
unit qualifying the setup as the wave faucet.
Under variation of the resonator length
multiple events of level crossing occur when the BSCs in the Friedrich-Wintgen scenario of wave localization
through the full destructive interference of decaying resonant
modes arise \cite{Friedrich,Volya,SBR}. For coaxial attachment this
scenario of the BSCs was considered in Ref. \cite{Lyapina} in
sector $m=0$. Similar effects take place in sectors $m\neq 0$ with
the BSCs degenerate with respect to the sign $\pm m$. Each BSC
with $m\neq 0$ gives rise to flows of acoustic intensity spinning
clockwise or anticlockwise inside the resonator. The importance of
such spinning trapped modes in acoustics was first shown by Parker
for axial flow compressor \cite{Parker}. Their existence in a
cylindrical acoustic infinitely long waveguide which contains rows
of large numbers of blades arranged around a central core, was
first reported by Duan and McIver \cite{Duan}. The BSCs have
become one of actively studied phenomena in different physical
areas \cite{Hsu review} because of potential applications in
lasing \cite{Kante,Zhang} and light enhancement in photonics
\cite{Yoon,Zhang&Zhang,Appl_Sc,Arxiv}.
Attachment of waveguides in non-coaxial manner breaks the
azimuthal symmetry mixing all subspaces  $m$ in wave transmission.
The most striking effect is that the evanescent modes of the waveguides lifts a degeneracy
of the eigenlevels with respect to the sign of $m$. Then for arbitrary
length of the resonator twist BSCs arise at selected angles
$\Delta\phi_c$. Because of absence of the azimuthal symmetry such
BSCs are not spinning.

\section{Wave faucet}
In this section we consider wave transmission in dependence on
angle $\Delta\phi$, the length of the resonator and frequency. For
numerical simulations we employ the acoustic coupled mode theory
which is in fact the effective non-Hermitian  Hamiltonian approach
\cite{Rotter,SR} adapted for the Neumann boundary conditions
\cite{Maksimov}. The propagating modes in sound hard cylindrical
waveguides are described by
\begin{eqnarray}\label{propag}
&\psi_{pq}(\rho,\alpha,z)=\phi_{pq}(\rho)\frac{1}{\pi
\sqrt{2k_{pq}}}
e^{ip\alpha+ik_{pq}z},&\\
&\phi_{0n}(\rho)=\frac{1}{J_0(\mu_{_{0}q})}J_0(\mu_{_{0}q}\rho),&\nonumber\\
&\phi_{pq}(\rho)=\sqrt{\frac{1}{\mu_{pq}^2-p^2}}\frac{\mu_{pq}}
{J_p(\mu_{pq})}J_p(\mu_{pq}\rho)&,\nonumber
\end{eqnarray}
where $\rho$, are $\alpha$ the polar coordinates,
$\mu_{pq}$ is the q-th root of equation $J_p'(\mu_{pq}\rho)|_{\rho=1}=0$,
\begin{equation}\label{kmn}
    k_{pq}^2=\omega^2-\mu_{pq}^2
\end{equation}
The quantities $\rho,~z, ~k_{pq}$ are dimensionless via the radius
of the waveguide. The frequency is measured in the units of the
ratio of the sound velocity to the radius of the waveguides. Sound
propagating bands are classified by two integers: the azimuthal
number $p=0, \pm 1, \pm 2, \ldots$ and $q=1, 2, 3, \ldots$ with
shapes of two degenerate transversal solutions $\cos p\alpha$ and
$\sin p\alpha$ depicted in Table \ref{Tab1}.
\begin{table}
\caption{Propagating bands and corresponding radial eigenmodes $Re(\phi_p(r,\phi))$}
\begin{tabular}{|c|c|c|c|c|}
  \hline
Channel &Bottom of band & Indexes & mode shapes \\
  \hline
1&0 & $p=0, q=1$ & \includegraphics[height=1cm,clip=]{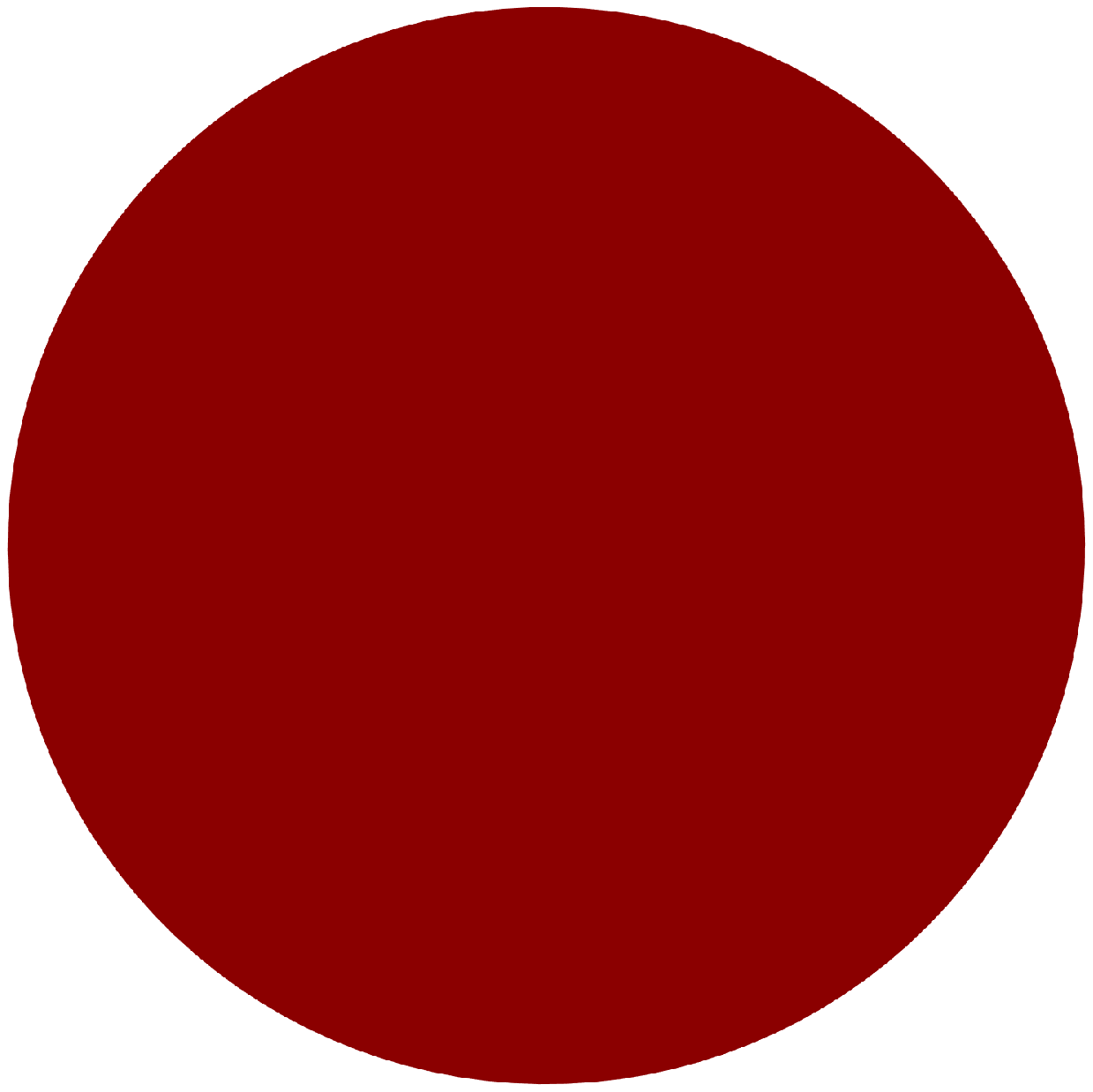} \\
  \hline
2&1.84118 & $p=\pm 1, q=1$ & \includegraphics[height=1cm,clip=]{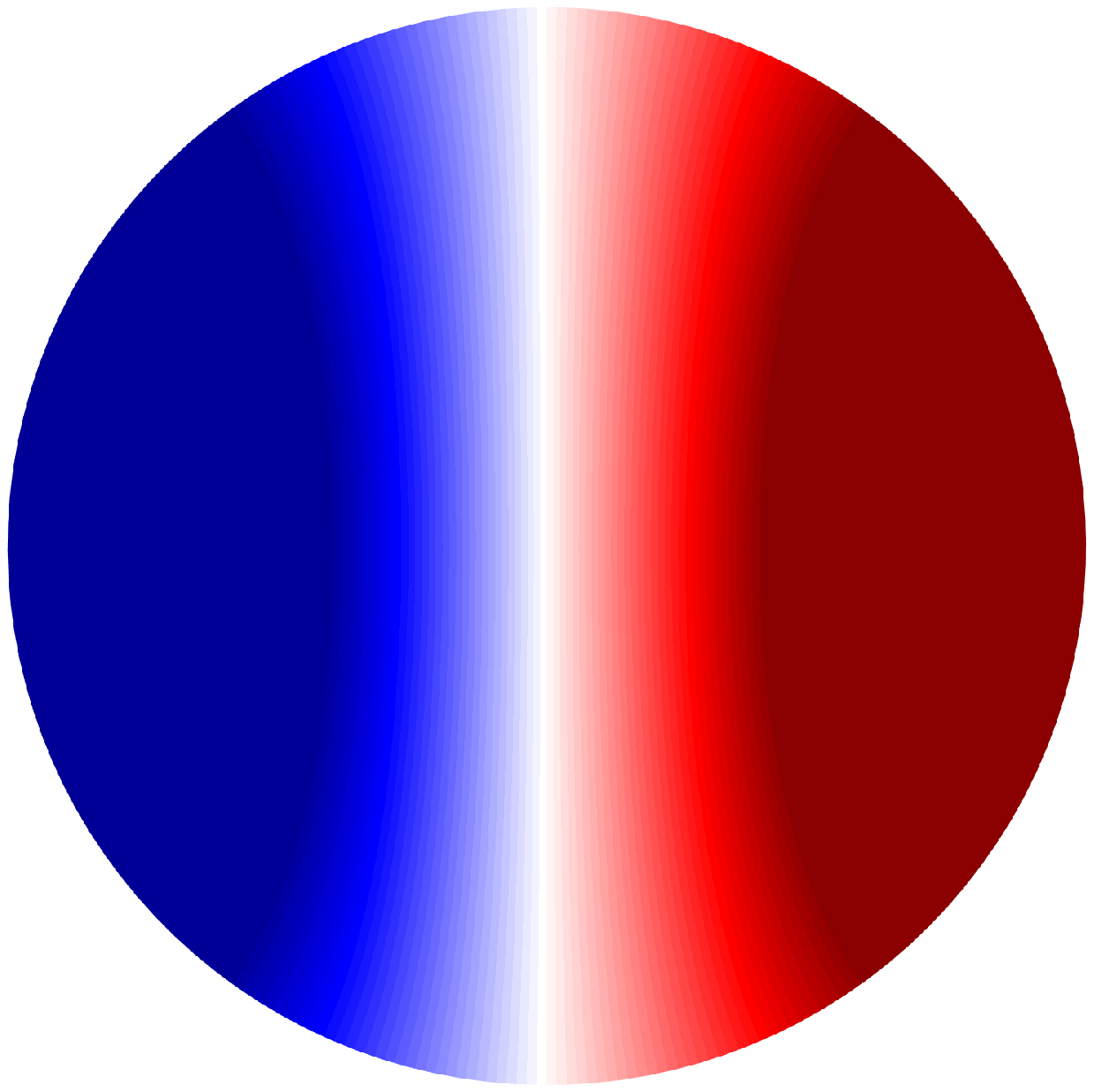} \\
  \hline
3&3.0542 & $p=\pm 2, q=1$ & \includegraphics[height=1cm,clip=]{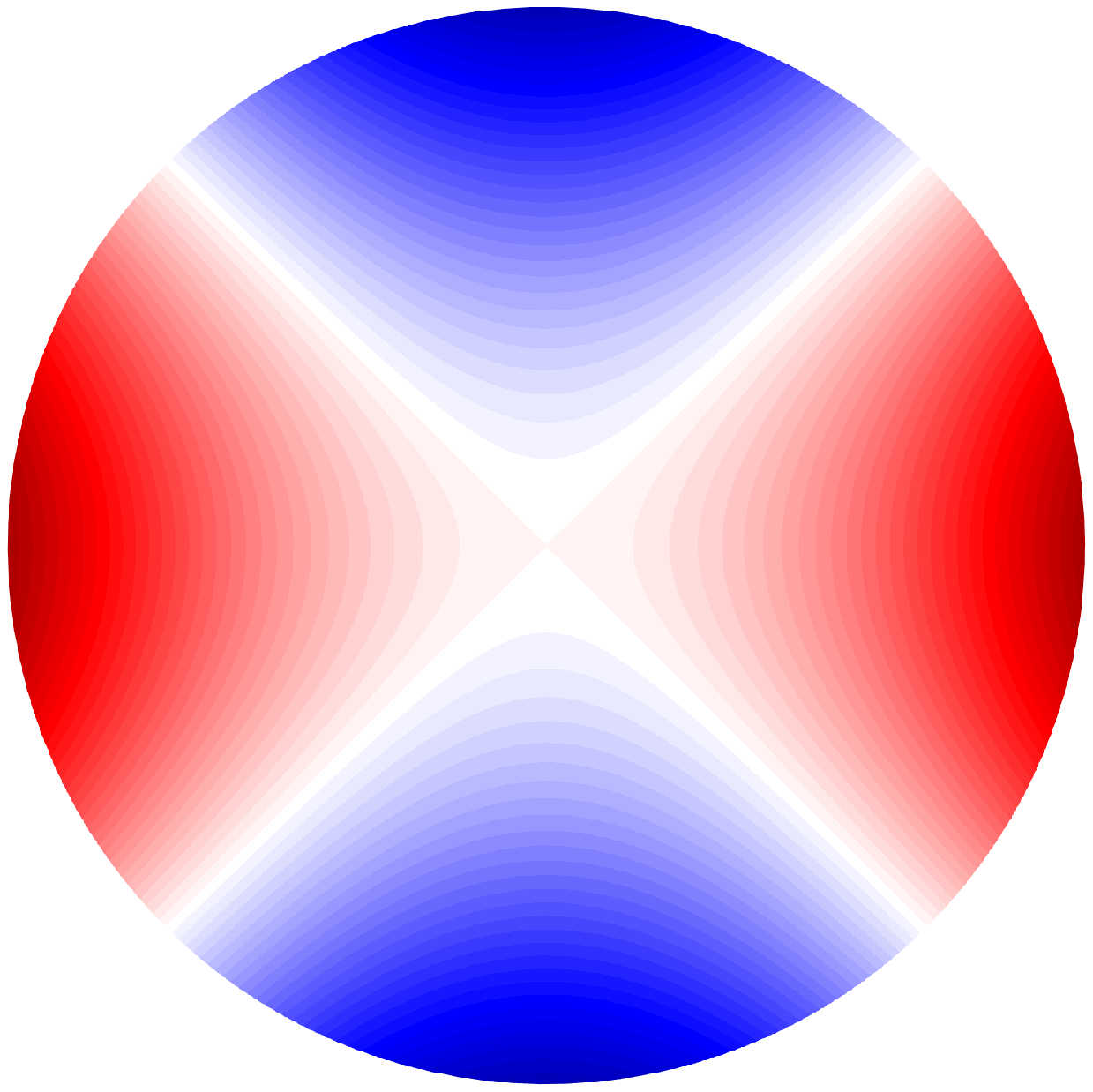} \\
  \hline
4&3.831706 & $p=0, q=2$ & \includegraphics[height=1cm,clip=]{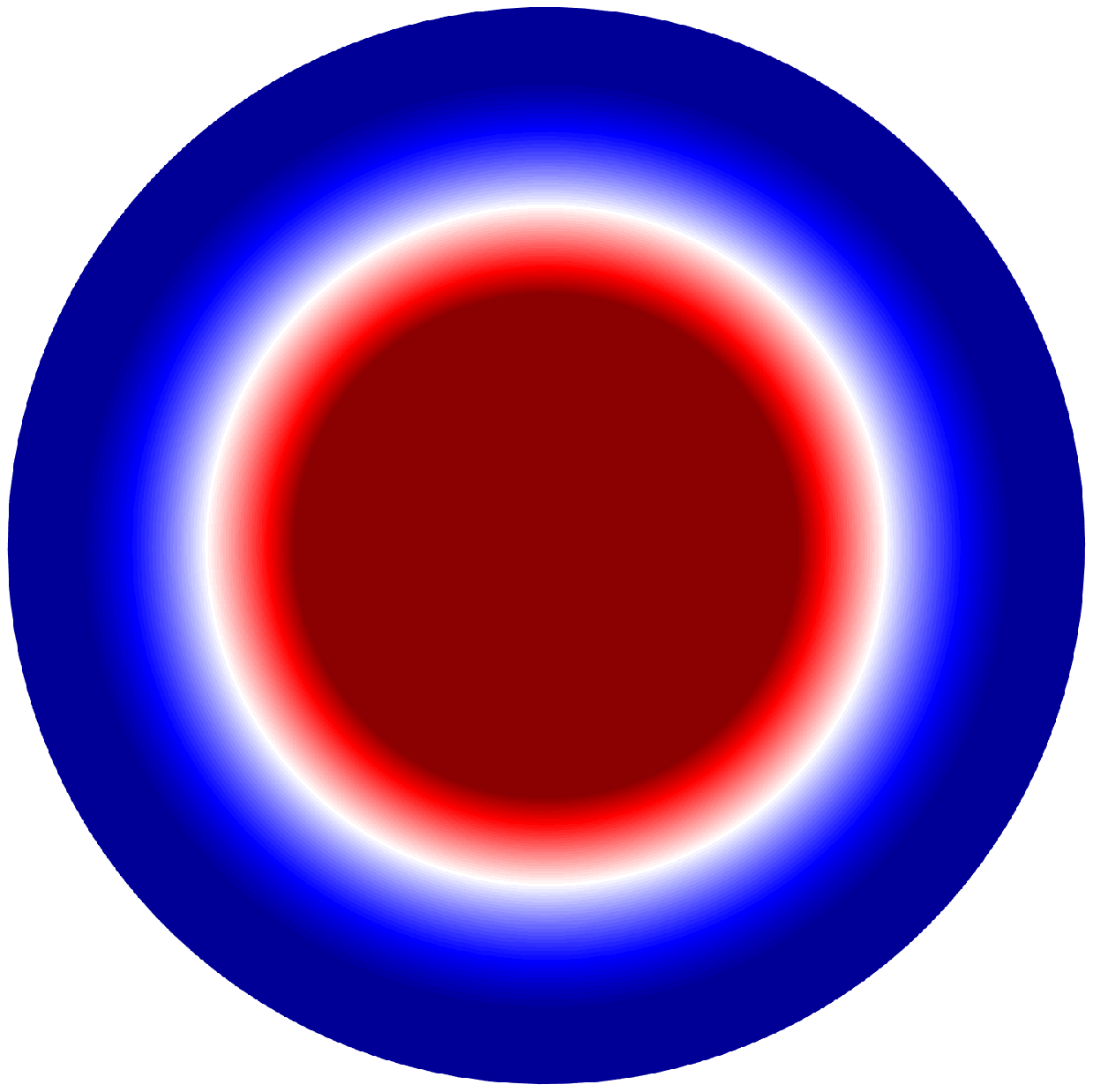} \\
  \hline
5&4.2012 & $p=\pm 3, q=1$ & \includegraphics[height=1cm,clip=]{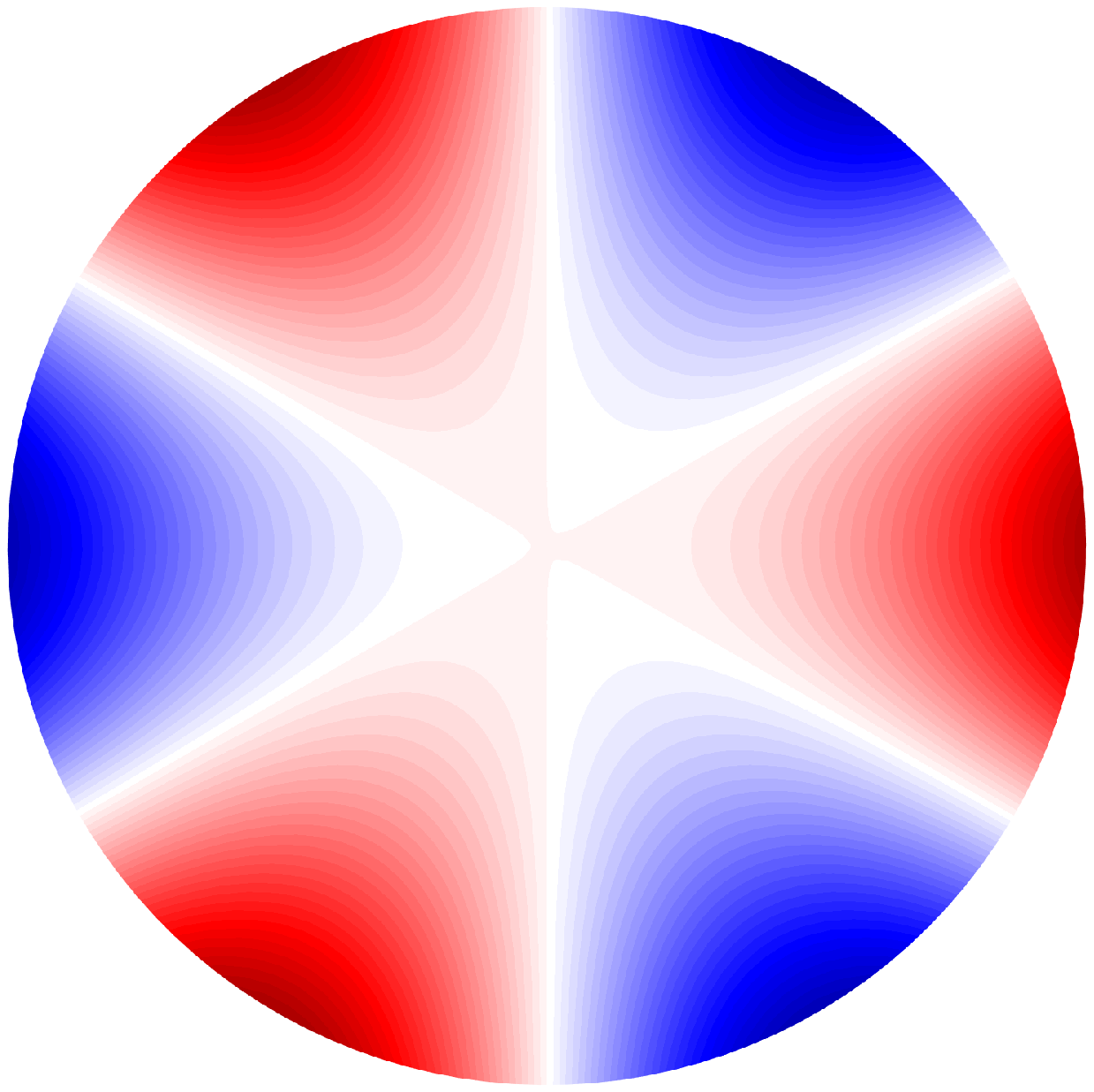} \\
  \hline
 \end{tabular}
 \label{Tab1}
\end{table}

The inner Hilbert space of the closed cylindrical resonator is
spanned by the eigenmodes
\begin{equation}\label{eigen}
\Psi_{mnl}(r,\phi,z)=\psi_{mn}(r)\sqrt{\frac{1}{2\pi}}\exp(im\phi)\psi_l(z),
\end{equation}
where
\begin{eqnarray}\label{eigRes}
&\psi_{0n}(r)=\frac{\sqrt{2}}{RJ_0(\mu_{_{0}n})}J_0(\mu_{_{0}n}r),&\nonumber\\
&\psi_{mn}(r)=\sqrt{\frac{2}{\mu_{mn}^2-m^2}}\frac{\mu_{mn}}
{RJ_m(\mu_{mn})}J_m(\mu_{mn}r/R), m\neq 0 ,&\nonumber\\
   & \psi_l(z)=\sqrt{\frac{2-\delta_{l,1}}{L}}\cos[\pi(l-1)z/L], l=1, 2, 3, \ldots &
\end{eqnarray}
$l=1, 2, 3, \ldots$ and $z$ is measured in terms of the waveguide radius.
The corresponding eigenfrequencies are
\begin{equation}\label{eig3D}
\omega_{mnl}^2=\left[\frac{\mu_{mn}^2}{R^2}+\frac{\pi^2 (l-1)^2}{L^2}\right]
\end{equation}
where $\mu_{mn}$ is the n-th root of equation following from the
Neumann boundary condition on the walls
$$J_m'(\mu_{mn}r)|_{r=R}=0$$.

\begin{figure}
\includegraphics[width=6cm,clip=]{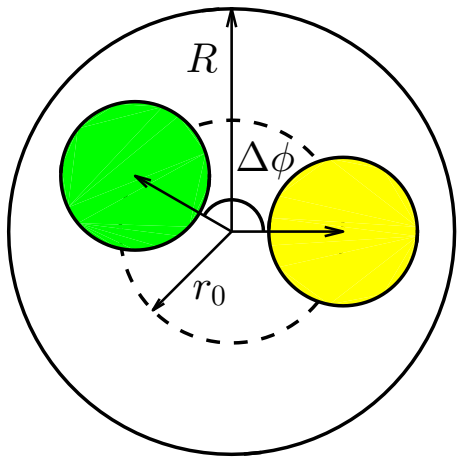}
\caption{(Color online)  Filled areas show overlapping integration area
in the coupling matrix (\ref{Wp}).}
\label{figg1}
\end{figure}

Let us write the effective non-Hermitian Hamiltonian as
\cite{Lyapina,Maksimov}
\begin{equation}\label{Heff}
\widehat{H}_{eff}=\widehat{H}_R-i\sum_{C=L,R}
\sum_{pq}k_{pq}\widehat{W}_{C,pq}\widehat{W}_{C,pq}^{\dagger}.
\end{equation}
where $\widehat{H}_R$ describes the closed resonator.
The matrix elements of $\widehat{W}$ are given by overlapping integrals \cite{Maksimov,Pichugin}
\begin{equation}\label{Wp}
W^C_{mnl;pq}=\psi_l(z_C)\int_0^{2\pi} d\alpha\int_0^1\rho d\rho\psi_{pq}(\rho,\alpha)
\Psi_{mn}^{*}(r(\rho,\alpha),\phi(\rho,\alpha))
\end{equation}
where $z_C=0,L$ are the position of the edges of the resonator
along the z-axis. According to Eq. (\ref{eigen}) we have
\begin{equation}\label{psil}
\psi_l(z=0)=\sqrt{\frac{2-\delta_{l,1}}{L}}, \psi_l(z=L)=\psi_l(0)(-1)^{l-1}.
\end{equation}
Integration is performed over circular cross section of the
attached waveguides as shown in Fig. \ref{figg1}. According to
Fig. \ref{figg1} one can link the polar coordinates of the
resonator with that of the waveguide $$r\sin\phi=\rho\sin\alpha,
r\cos\phi=r_0+\rho\cos\alpha$$ where $r_0$ is the distance between
the axes of the waveguide and resonator.

Although the waveguides are identical they are attached to the
resonator at different azimuthal angles as shown in Fig.
\ref{figg1} to give rise to an exact relation between the coupling
matrix elements
\begin{equation}\label{WLWR}
W^R_{mnl;pq}=(-1)^{l-1}e^{im\Delta\phi}W_{mnl;pq}.
\end{equation}
Here $W_{mnl;pq}=W^L_{mnl;pq}$ are the coupling matrix elements of
the resonator modes specified by integers $m,n,l$ with $p,q$
propagating modes of the left waveguide (see Fig. \ref{fig1}).
Then the effective Hamiltonian takes the following form
\begin{eqnarray}\label{Heffp02}
&\langle mnl|\widehat{H}_{eff}|m'n'l'\rangle=
\omega_{mnl}^2\delta_{mm'}\delta_{nn'}\delta_{ll'}&\nonumber\\
&-i\sum_{pq}k_{pq}[1+(-1)^{l+l'}e^{i(m-m')\Delta\phi}]W_{mnl;pq}W^{*}_{m'n'l';pq}.&
\end{eqnarray}
The transmittance of sound waves through the resonator is given by
the inverse of the matrix $\widehat{G}=\widehat{H}_{eff}-\omega^2$
\cite{Rotter,Maksimov}
\begin{equation}\label{S-matrix}
T_{pq;p'q'}=-2ik_{pq} \sum_{mnl}\sum_{m'n'l'}W_{mnl;pq}\langle
mnl|[\widehat{H}_{eff}-\omega^2]^{-1}
e^{i(m-m')\Delta\phi}|m'n'l'\rangle W^{*}_{m'n'l';p'q'}.
\end{equation}

\begin{figure}
\includegraphics[width=8cm,clip=]{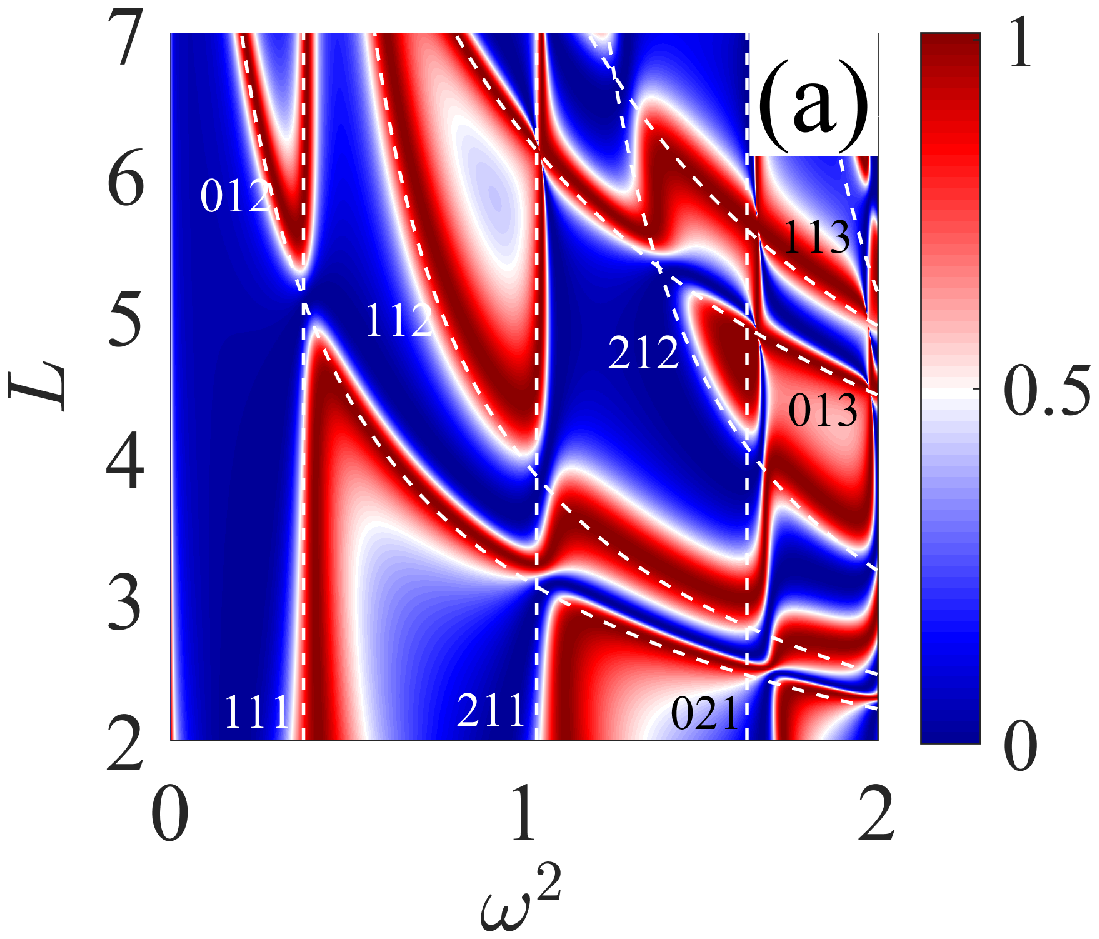}
\includegraphics[width=8cm,clip=]{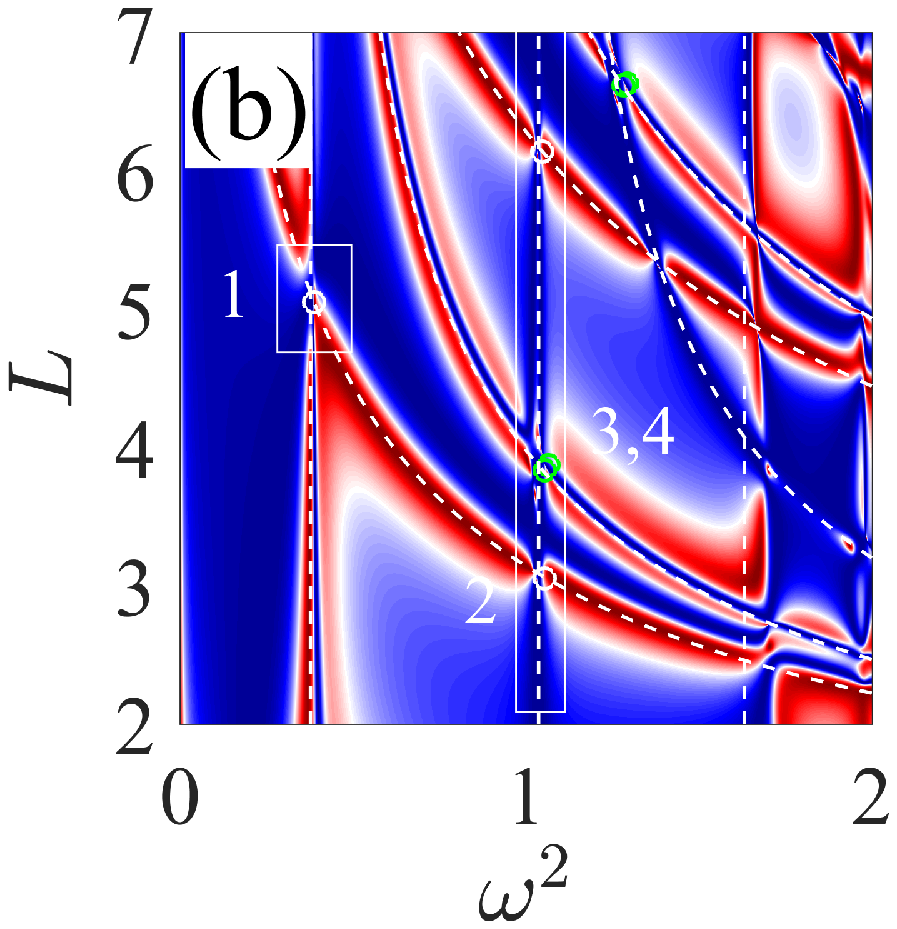}
\includegraphics[width=8cm,clip=]{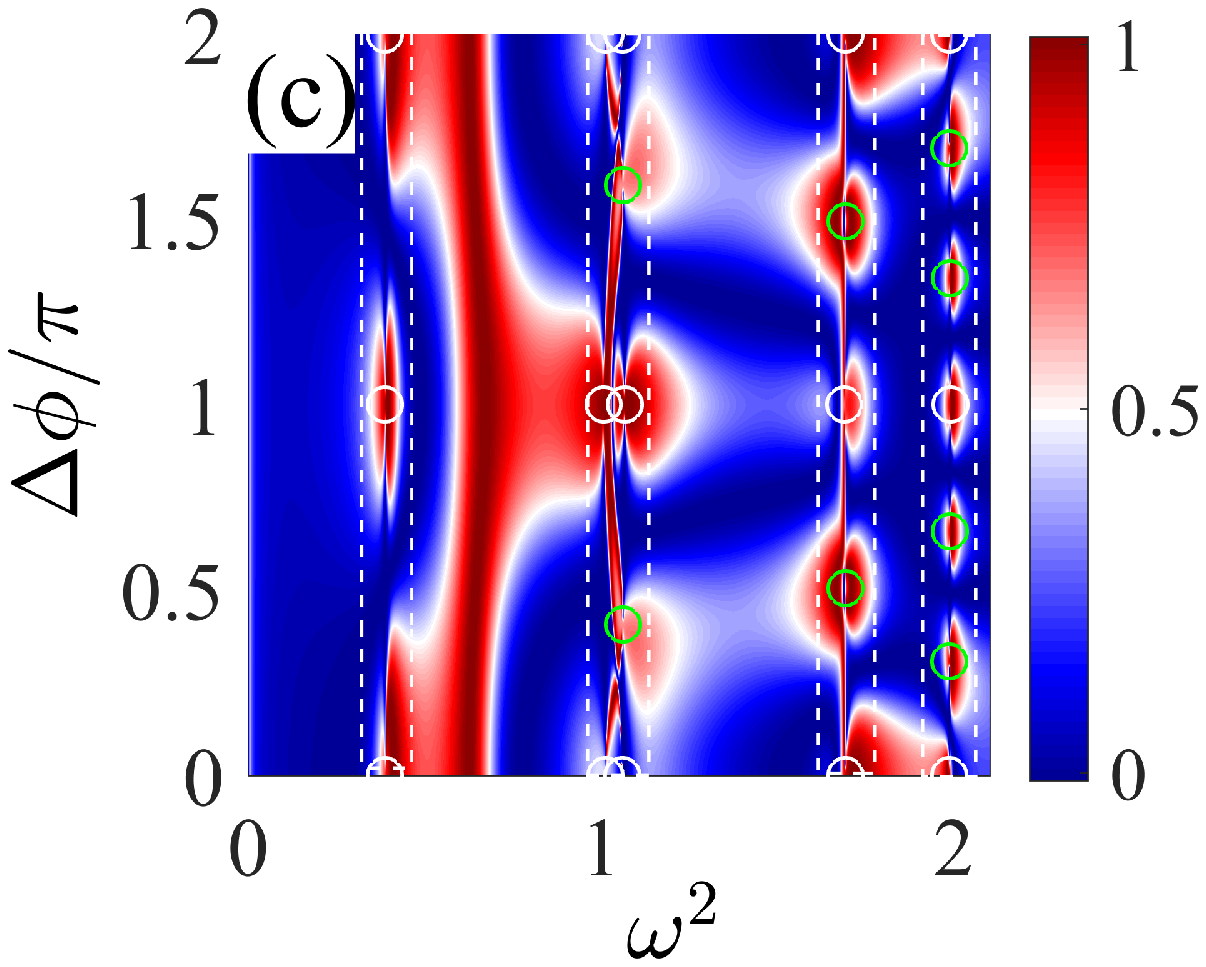}
\includegraphics[width=8cm,clip=]{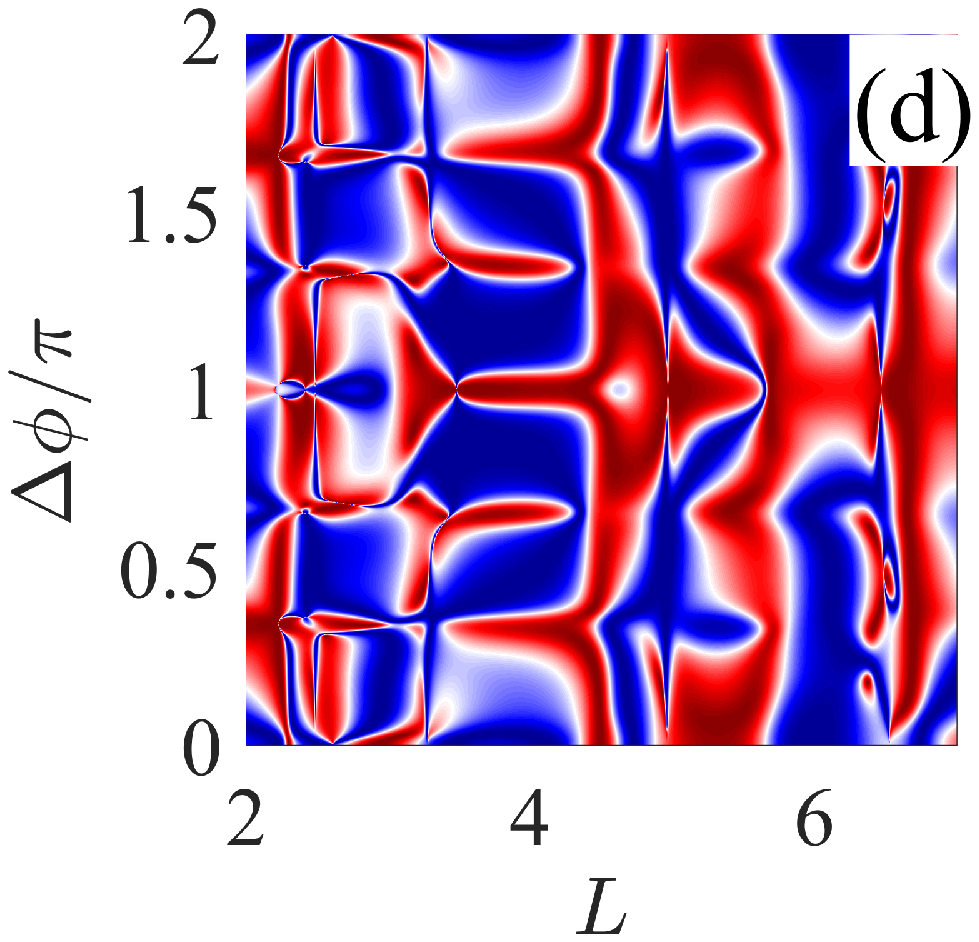}
\caption{(Color  online)  Transmittance  of a  cylindrical
resonator vs frequency and length  of the  resonator $L$ at (a)
$\Delta\phi=0$ and (b) $\Delta\phi=\pi/4$. (c) vs frequency and
rotation angle $\Delta\phi$ at $L=4$ and (d) vs length and
rotation angle at $\omega^2=2$. Dash lines in (a) show
eigenfrequencies of closed resonator with corresponding indexes $m
n l$. The positions of  the BSCs are shown by closed circles.}
\label{fig2}
\end{figure}

In what follows we take both waveguides with unit radius shifted
relative to the central axis of the resonator with the radius
$R=3$ by a distance $r_0=1.5$. We consider transmission in the
first channel $p=0, q=1$ in the frequency domain $0< \omega <
1.8412$ as shown in Table \ref{Tab1}. The transmittance versus the
squared frequency and the resonator length $L$
shown in Fig. \ref{fig2} (a) and (b) for $\Delta\phi=0, \pi/4$ respectively. The resonant behavior
of the transmittance is due to that the resonator is three-dimensional.
The coupling matrix elements
(\ref{Wp}) can be estimated as the ratio of the cross-section of
the waveguides to the cross-section of the resonator. The resonant
widths are proportional to squared coupling matrix elements, i.e.,
proportional to $R^{-4}$, while the distance between the
eigenlevels is proportional to $\Delta E \sim R^{-2}$. Hence for
$R=3$ we have the regime of weak coupling \cite{Sok&Zel} while for
two-dimensional resonator one would have $\Gamma \sim R^{-2}$ as
well as $\Delta E \sim R^{-2}$, i.e., the regime of overlapping
resonances \cite{Lyapina}. In the two-dimensional systems the
regime of weak coupling can be reached only by use of special
diaphragms \cite{Kuhl}. Indeed Fig. \ref{fig2} (a) and (b) shows
that the transmittance of the resonator basically follows the
eigenvalues of the closed resonator.

\begin{figure}
\includegraphics[width=8cm,clip=]{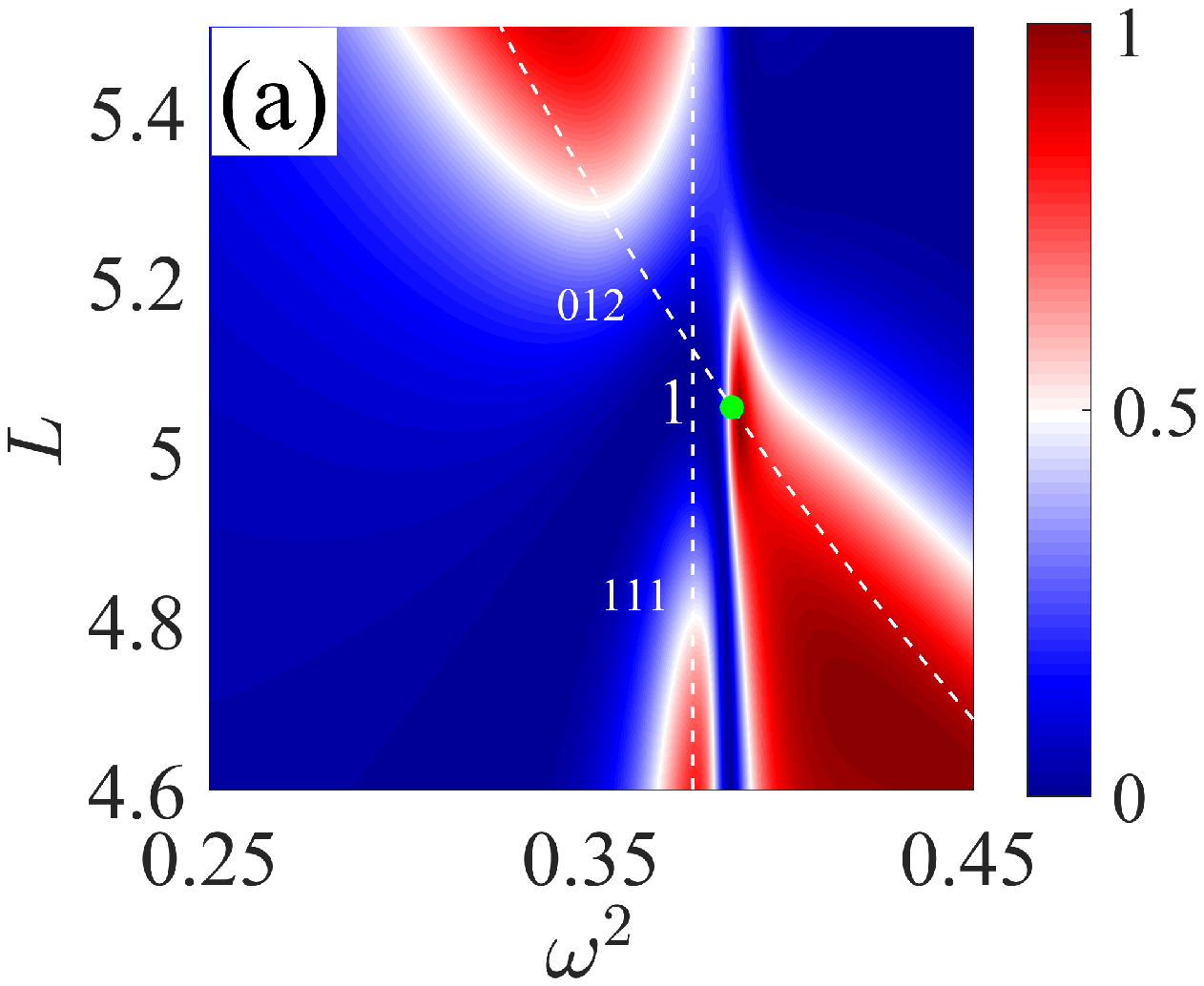}
\includegraphics[width=8cm,clip=]{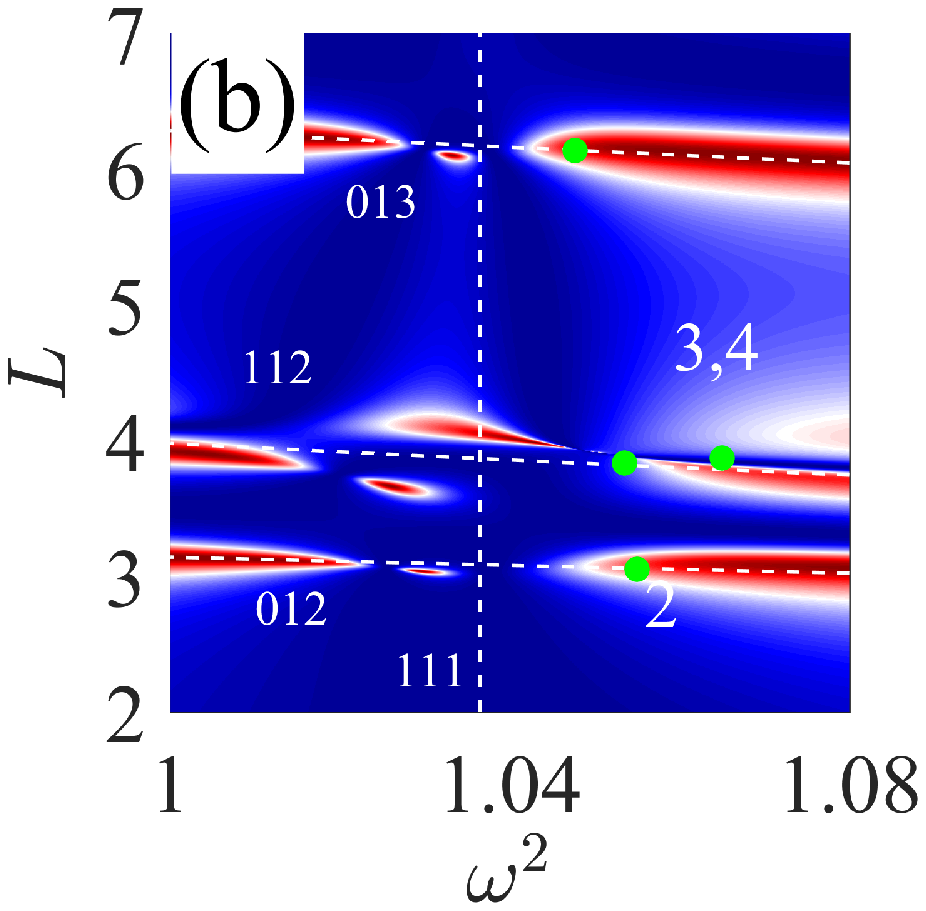}
\caption{(Color  online)  Transmittance  vs frequency and length at $\Delta\phi=\pi/4$ at $L=4$
in the domains shown  in Fig. \ref{fig2} (b) by  rectangles.}
\label{fig2bfine}
\end{figure}

\begin{figure}
\includegraphics[width=6cm,clip=]{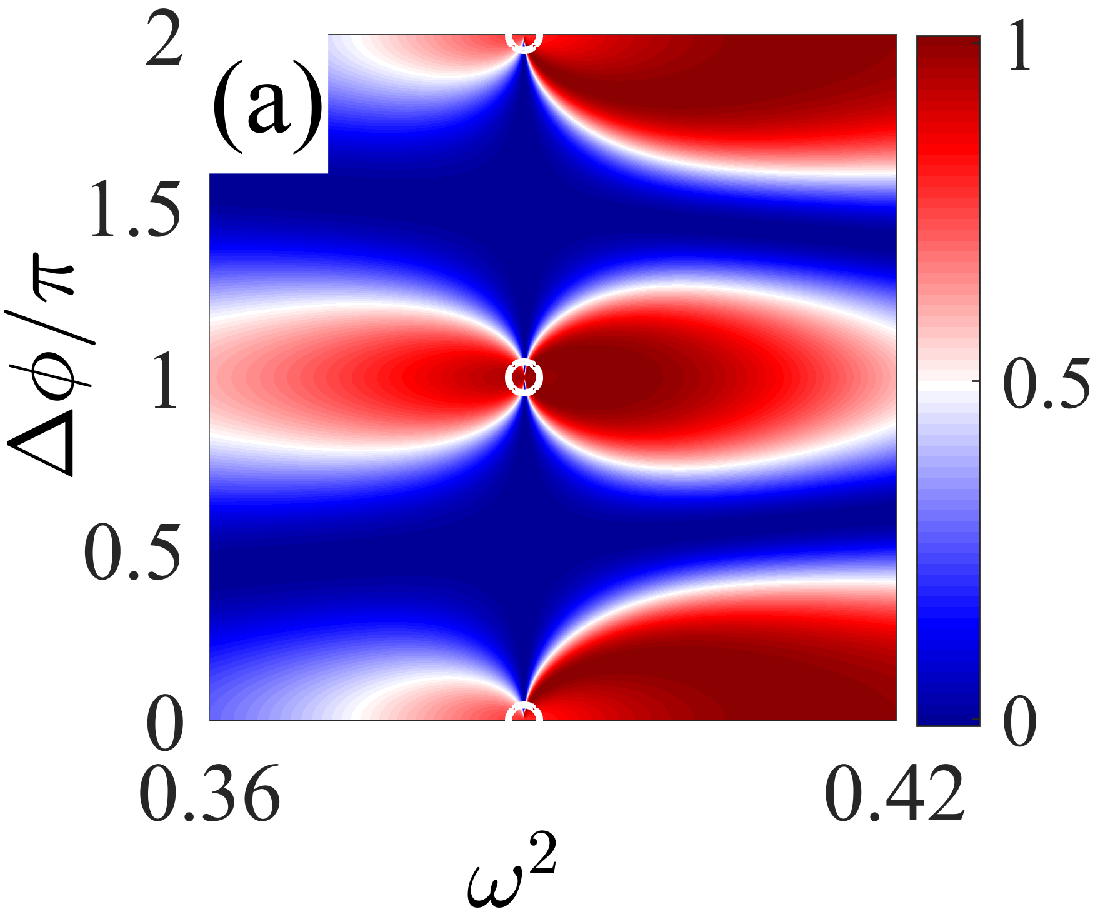}
\includegraphics[width=6cm,clip=]{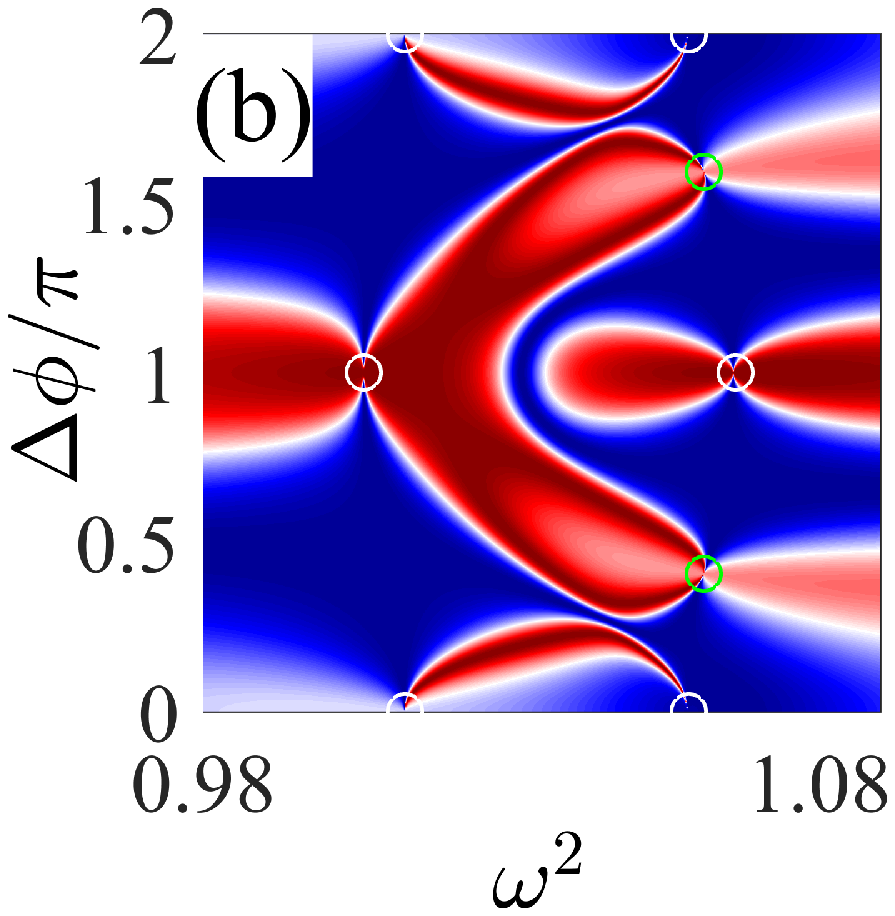}
\includegraphics[width=6cm,clip=]{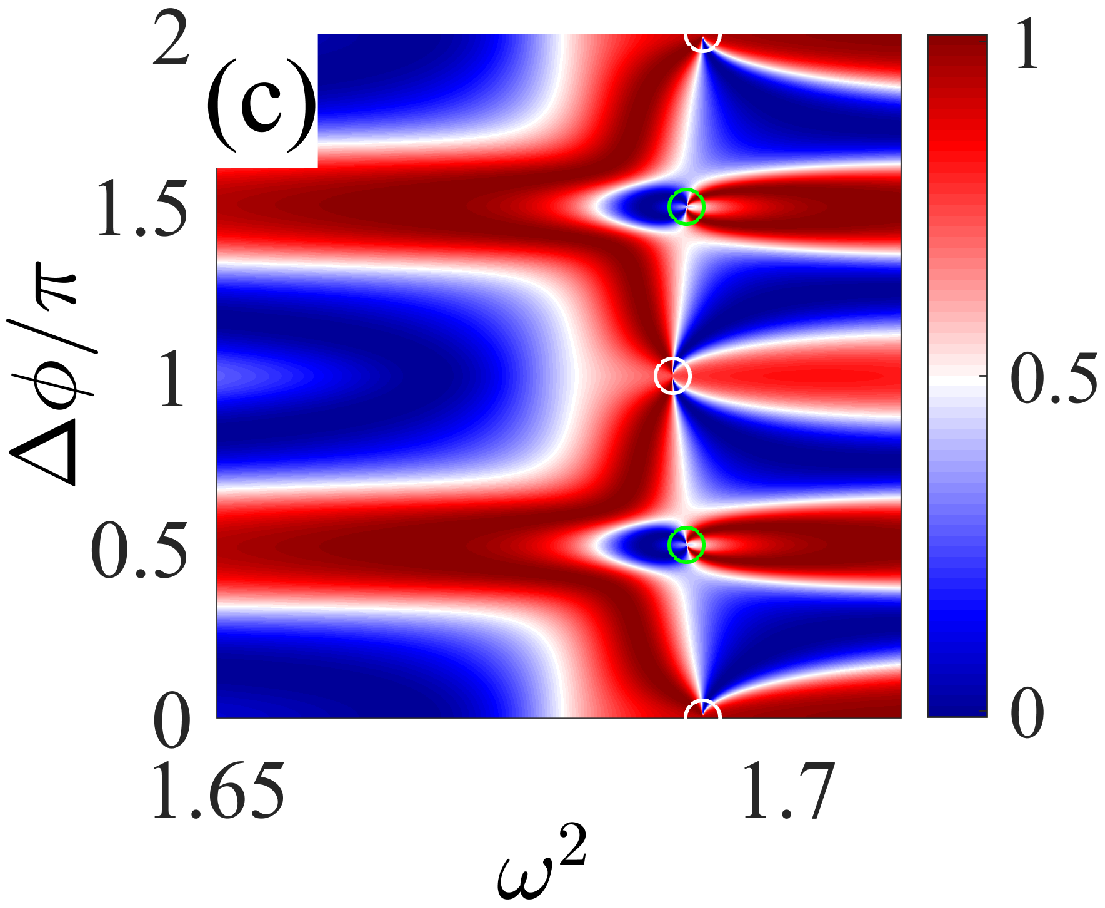}
\includegraphics[width=6cm,clip=]{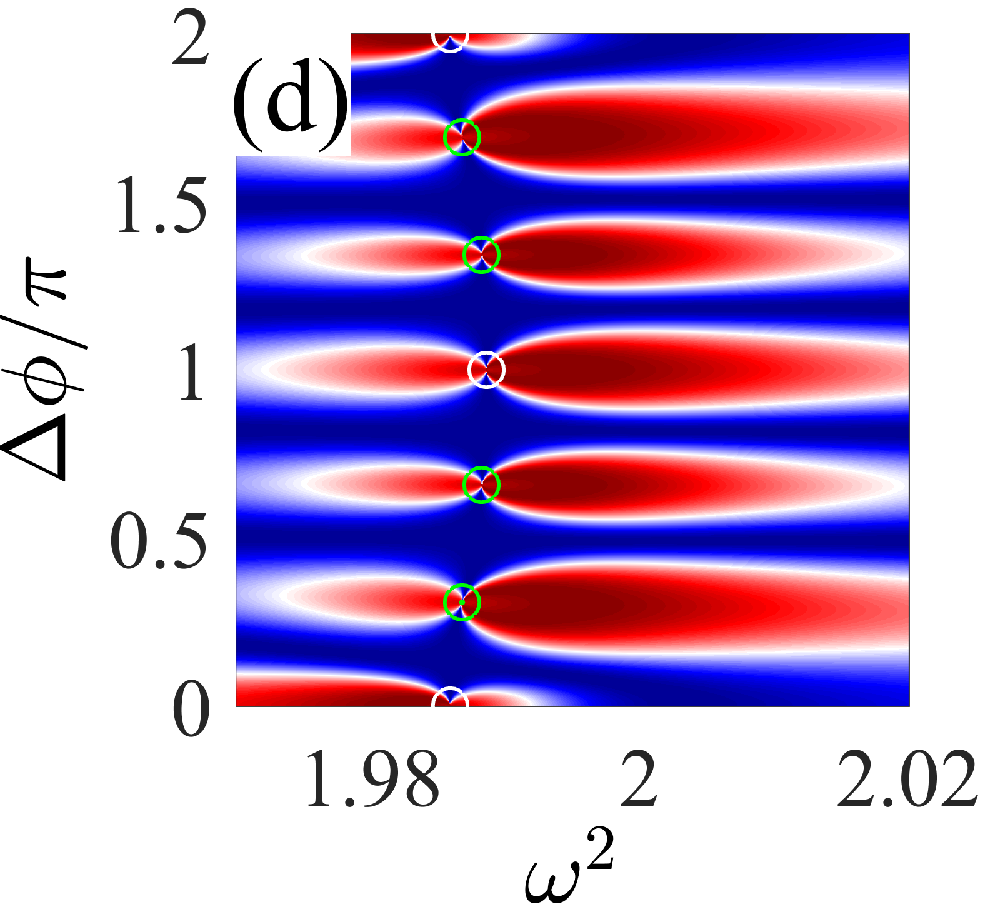}
\caption{(Color  online)  Transmittance  vs frequency and rotation angle $\Delta\phi$ at $L=4$
in frequency domains shown  in Fig. \ref{fig2} (c) by  dash rectangles.  The  positions  of  the BSCs are shown by closed circles.}
\label{fig3}
\end{figure}

When $\Delta\phi\neq 0$ the transmittance undergoes changes in the
vicinity of eigenvalue crossings as shown in Fig. \ref{fig2} (b).
Figs. \ref{fig2} (c) and \ref{fig2} (d) brightly demonstrate the
effect of wave faucet when the resonator is opened or closed by
rotation of the waveguide.
The features of wave interference under rotation of the waveguide
are demonstrated in Figs. \ref{fig2bfine}, \ref{fig3}, and \ref{fig4} where the
transmittance is shown in higher resolution in zoomed windows.
First of all one can see a high sensitivity of the transmittance
to the rotation angle $\Delta\phi$. Moreover there are a mass of
points in the vicinity of which even extremely small changes in
parameters of the system cause drastic changes in the
transmittance. These points are marked by open circles and
correspond to the BSCs. In the vicinity of the BSC points the unit
transmittance coalesces with zero transmittance (collapse of the
Fano resonance \cite{Kim,SBR}). In particular, one can see in Fig.
\ref{fig3} that even the slightest rotation opens and closes the
resonator illustrating the effects of wave faucet. These resonant
angular features are the result of evanescent modes and will be
studied analytically by use of truncated effective Hamiltonian.

\section{Bound states in the continuum}

\begin{figure}[ht]
\includegraphics[width=6cm,clip=]{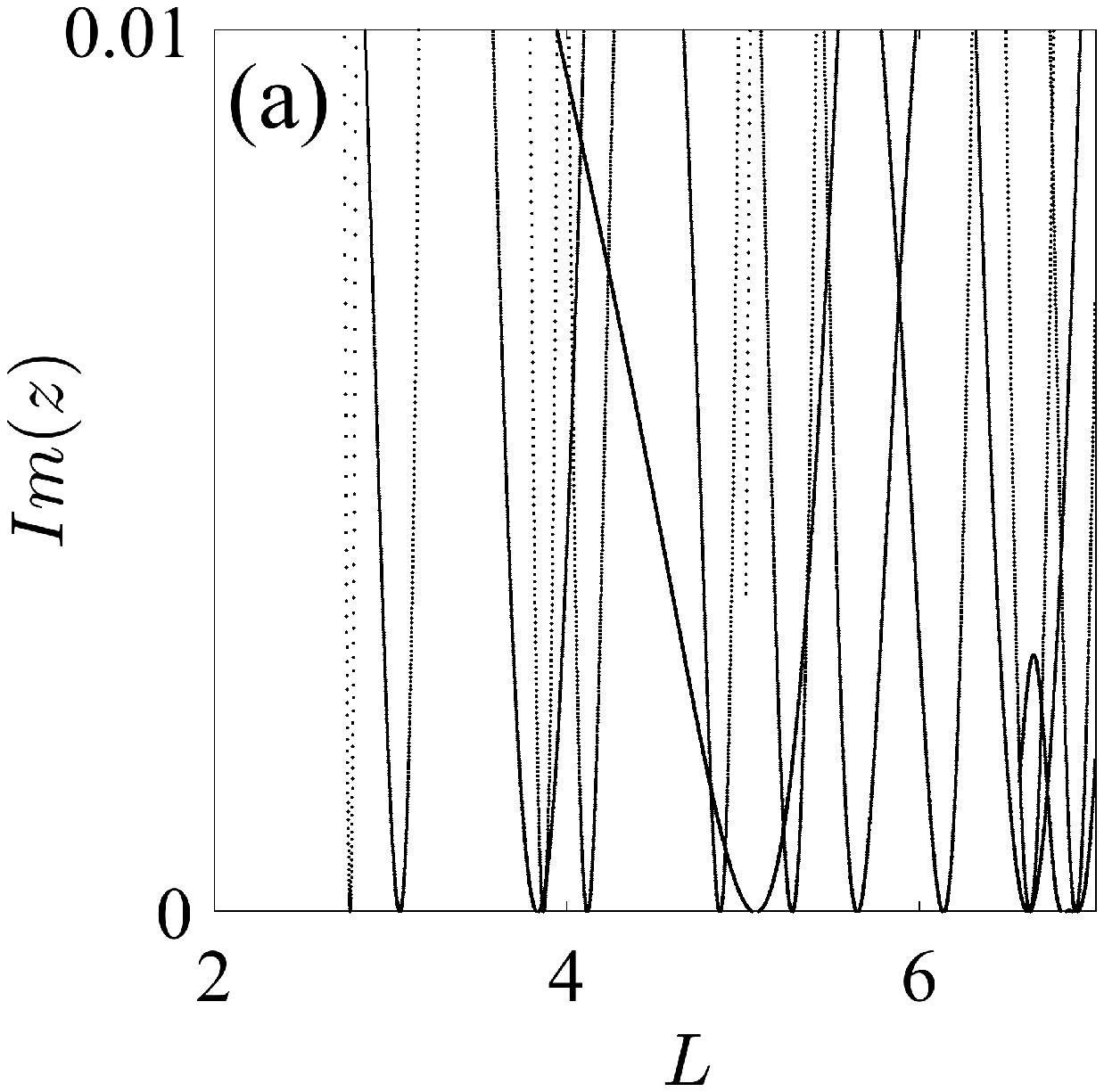}
\includegraphics[width=6cm,clip=]{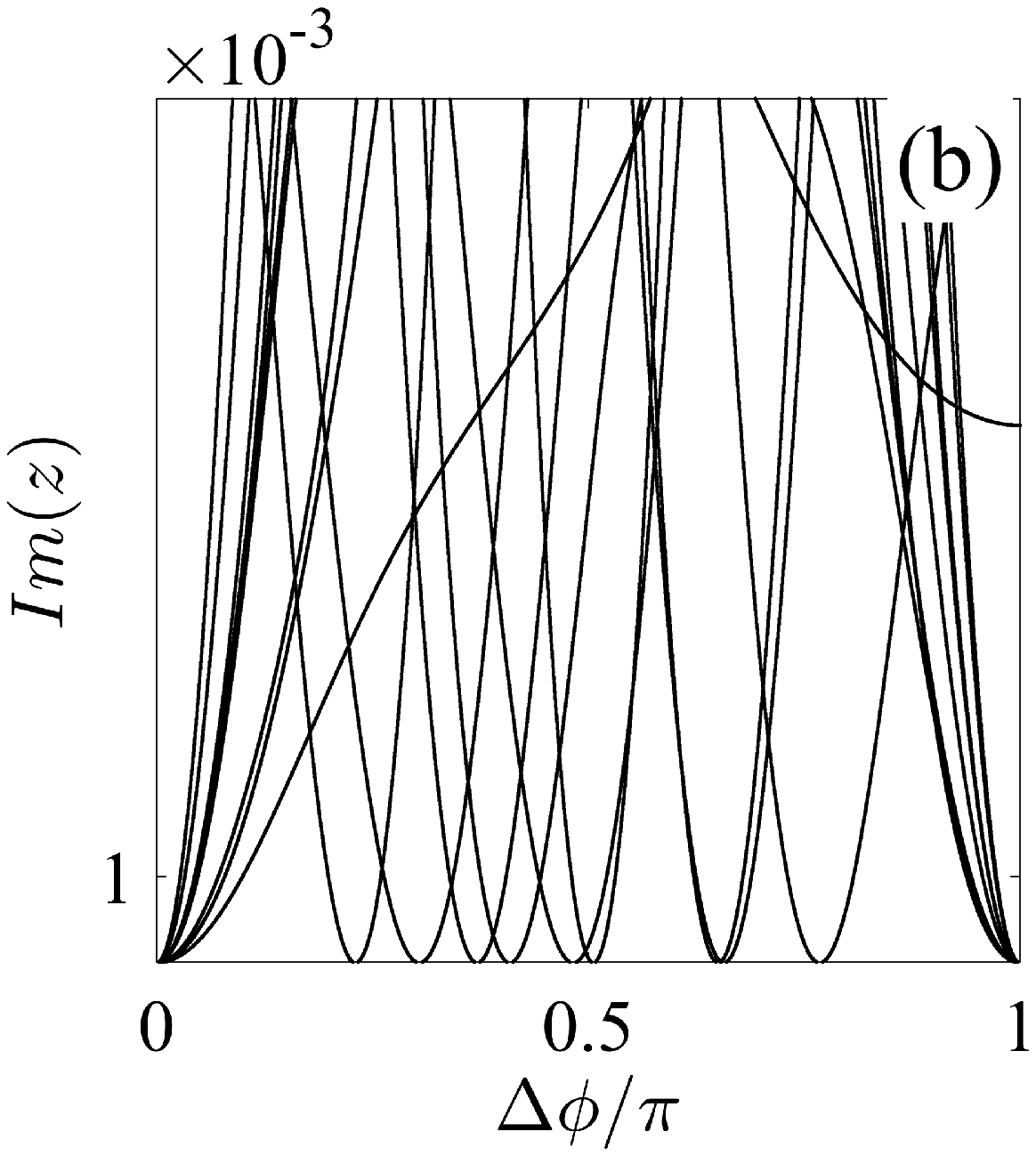}
\caption{(Colour  online) Evolution of resonant widths under
variation of (a) the resonator length at $\Delta\phi=\pi/4$
and (b) rotation of the waveguide at $L=4$.} \label{fig4}
\end{figure}

\begin{figure}[ht]
\includegraphics[width=6cm,clip=]{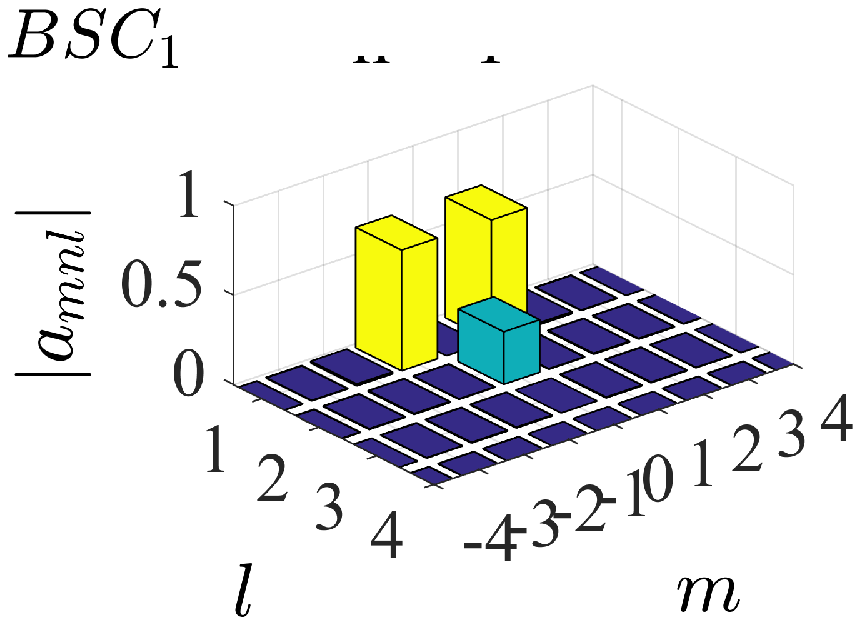}
\includegraphics[width=6cm,clip=]{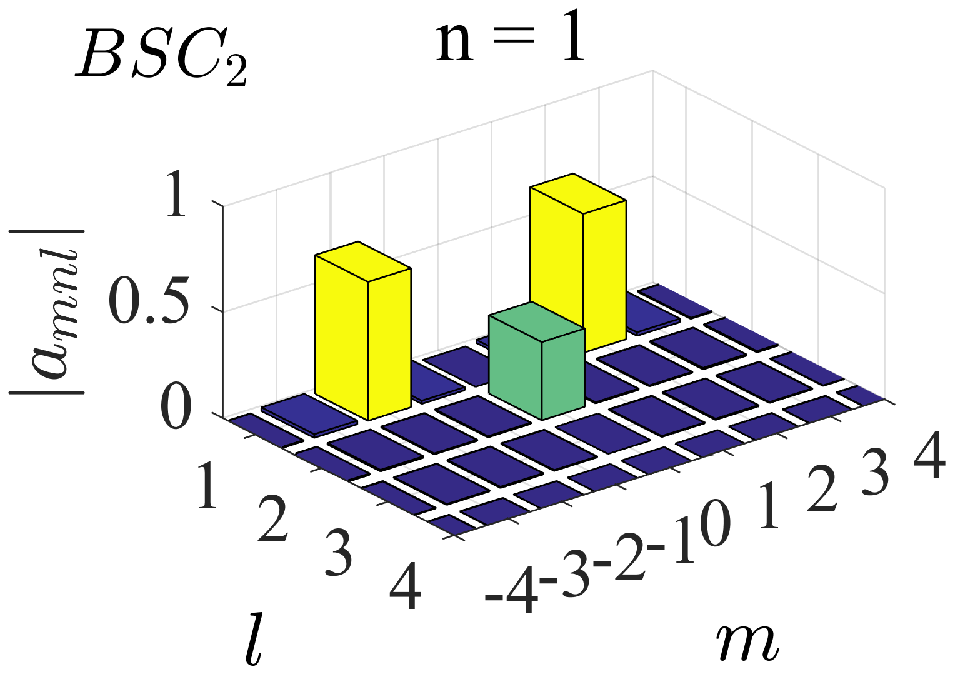}
\includegraphics[width=6cm,clip=]{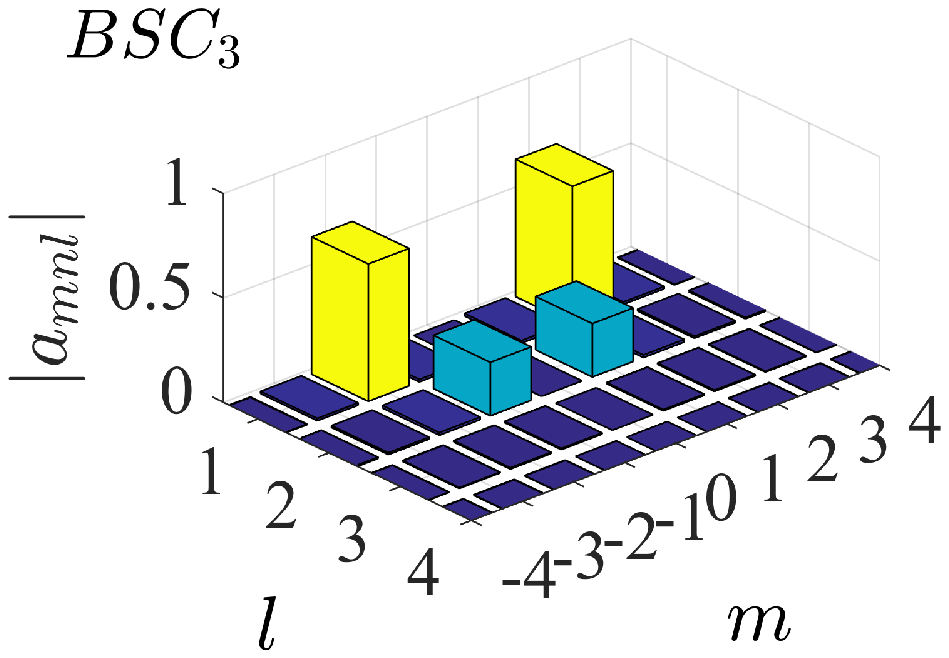}
\includegraphics[width=6cm,clip=]{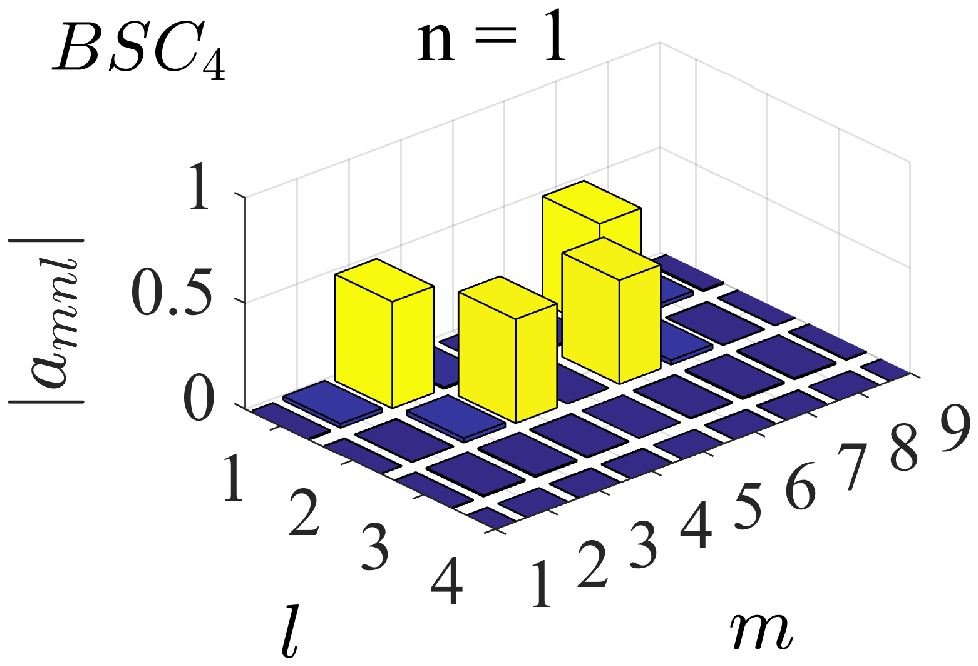}
\caption{(Color  online) Modal expansion coefficients $|a_{mnl}|$
of BSCs shown in Fig. \ref{fig2} for the case $\Delta\phi=\pi/4$.}
\label{fig5}
\end{figure}

In Figs. \ref{fig2}, \ref{fig2bfine} and \ref{fig3} we show BSC points by closed
and open circles. The procedure to find BSCs is based on search of
real eigenvalues of the effective non-Hermitian Hamiltonian
(\ref{Heff}) $z_{\lambda}(\omega,L,\Delta\phi)$ \cite{RS2005}. A
typical behavior of the imaginary parts (resonant widths) as
dependent on the length $L$ (a) and rotation angle $\Delta\phi$
(b) is shown in Fig. \ref{fig4}.
One can see numerous events of turning of resonant widths to zero.
Each BSC point is searched by solving the fix point equations
\cite{Rotter} $\omega=Re(z(\omega,L,\Delta\phi)),
~Im(z(\omega,L,\Delta\phi))=0$. After the fix point equation is
solved  we can determine the eigenmodes of the effective
Hamiltonian with real eigenvalues which are BSC mode shapes
\cite{Lyapina,SBR}. Fig. \ref{fig5} shows the expansion
coefficients $a_{mln}$ of the BSCs over the eigenmodes of the
closed resonator (\ref{eigen})
\begin{equation}\label{BSC expansion}
    \psi_{BSC}(r,\phi,z)=\sum_{mnl}a_{mnl}\Psi_{mnl}(r,\phi,z).
\end{equation}
The expansion coefficients $a_{mnl}$ are listed in Table
\ref{Tab2} for the BSCs enumerated from one to four in Fig.
\ref{fig2} (a).
\begin{table}
\caption{BSC points and their expansion coefficients $a_{mnl}$.}
\begin{tabular}{|c|c|c|c|c|c|}
  \hline
BSC number & $\omega^2$ & $L$ & $\Delta\phi$ & $mnl$& $a_{mnl}$ \\
  \hline
 &       &       &         & 012 & -0.113+0.272i\\
1& 0.385 & 5.065 & $\pi/4$ & 111 &  -0.478(1-i) \\
 &       &       &         & -111& 0.675\\
  \hline
 &       &       &         & 012 & -0.261(1-i)\\
2& 1.055 & 3.051 & $\pi/4$ & 211 &  0.656i \\
 &       &       &         &-211 &0.656\\
\hline
 &       &       &         & 211 & 0.658i\\
3& 1.0535& 3.833 & $\pi/4$ & -211 &  0.658 \\
 &       &       &         &112 &-0.237-0.098i\\
 &       &       &         &-112 &-0.098-0.237i\\
\hline
 &       &       &         & 211 & -0.505\\
4& 1.065& 3.869 & $\pi/4$ & -211 &  0.505 \\
 &       &       &         &112 &-0.455-0.189\\
 &       &       &         &-112 &0.189+0.455i\\
\hline
 \end{tabular}
 \label{Tab2}
\end{table}

From Fig. \ref{fig2} (a), (b) and Fig. \ref{fig2bfine} one can see that the BSCs are
positioned in the vicinity of  degeneracy points. However one can
see from Fig. \ref{fig2bfine} (a) that the BSC 1 are not spaced
exactly at the degeneracy point of eigenlevels $012$ and $\pm 111$
due to the evanescent modes \cite{SBR}.  The second feature of the
BSCs is the frequency splitting in the vicinity of the crossing of
eigenlevels with $m=\pm 1$ and $m=\pm 2$. Such split  BSCs 3 and 4
are shown by closed circles in Fig. \ref{fig2bfine} (b).
Fig. \ref{fig5} shows that the BSCs are composed of the eigenmodes
of the closed resonator $\Psi_{mnl}(r,\phi,z)$ which undergo
degeneracy in the vicinity of the BSC point. Surprisingly,
although BSCs are superposed of complex functions $e^{im\phi}$
they do not support flows of acoustic intensity in contrast to the
case of coaxial attachment of waveguides \cite{Duan}. As Fig.
\ref{fig5} shows each pair of eigenmodes with $\pm m$ contributes
into the BSC with equal $|a_{mnl}|$.

\section{BSCs in two continua different in phase}
Displacement of the waveguide relative to the resonator does not
change its continuous spectrum. However the coupling matrix
elements of the resonator eigenmodes with the continua are subject
to alternation to affect the transmission. In particular under
rotation of one of the waveguides the matrix elements acquire
phase shift (\ref{WLWR}). Therefore in the framework of the
effective non-Hermitian Hamiltonian  one can say that two continua
become different by phase. First, the problem of the BSC residing
in a finite number of continua was considered by Pavlov-Verevkin
and coauthors \cite{Pavlov}. Rigorous statement about the BSCs was
formulated as follows. The interference among $N$ degenerate
states which decay into K non-interacting continua generally leads
to the formation of $N-K$ BSCs. The equivalent point of view
\cite{SBR,JPC28} is that the linear superposition of the $N$
degenerate eigenstates $\sum_{n=1}^N a_n\psi_n$ can be adjusted to
have zero coupling with $K$ different continua in $N-K$ ways by
variation of the $N$ superposition coefficients $a_n$.
Respectively, these coefficients $a_n$ define an expansion of the
BSC over the eigenstates of the closed resonator. The number of
continua can grow due to a number of reasons, for example, non-
symmetrically attached waveguides, multiple propagation subbands
in the waveguides, or two polarizations of the radiation continuum
in case of electromagnetic BSCs. Each case puts the problem of
constructing BSCs in the case of many continua on the line of art
\cite{Appl_Sc,JPC28,anneal,Wei,Bo Zhen,Yang}.

In the present case of waveguide rotation  the number of continua
has been doubled for the frequency of sound waves below the second
propagation threshold $\omega<\mu_{11}$ (see Table \ref{Tab1}).
Therefore we could expect BSCs only at the points of threefold
degeneracy in compliance with the generic statement \cite{Pavlov}.
There are indeed numerous points where the eigenlevels
$\omega_{mnl}^2$ double degenerate in $\pm m$ cross the
eigenlevels $\omega_{0nl}^2$ with BSCs marked by open circles in
Fig. \ref{fig2} (b) and (c). However the zoomed picture of
transmission in Fig. \ref{fig2bfine} (a) shows that this
conclusion is only approximate. Thus, the case needs in special
consideration. As shown in Table \ref{Tab2} the BSCs are
superposed from only a few eigenmodes. Hence we can truncate the
effective Hamiltonian to the relevant eigenmodes similar to that
in Refs. \cite{Friedrich,Volya,SBR}.

\subsection{The mode with $m=0$ crosses the modes with $\pm m$}
\begin{figure}[ht]
\includegraphics[width=6cm,clip=]{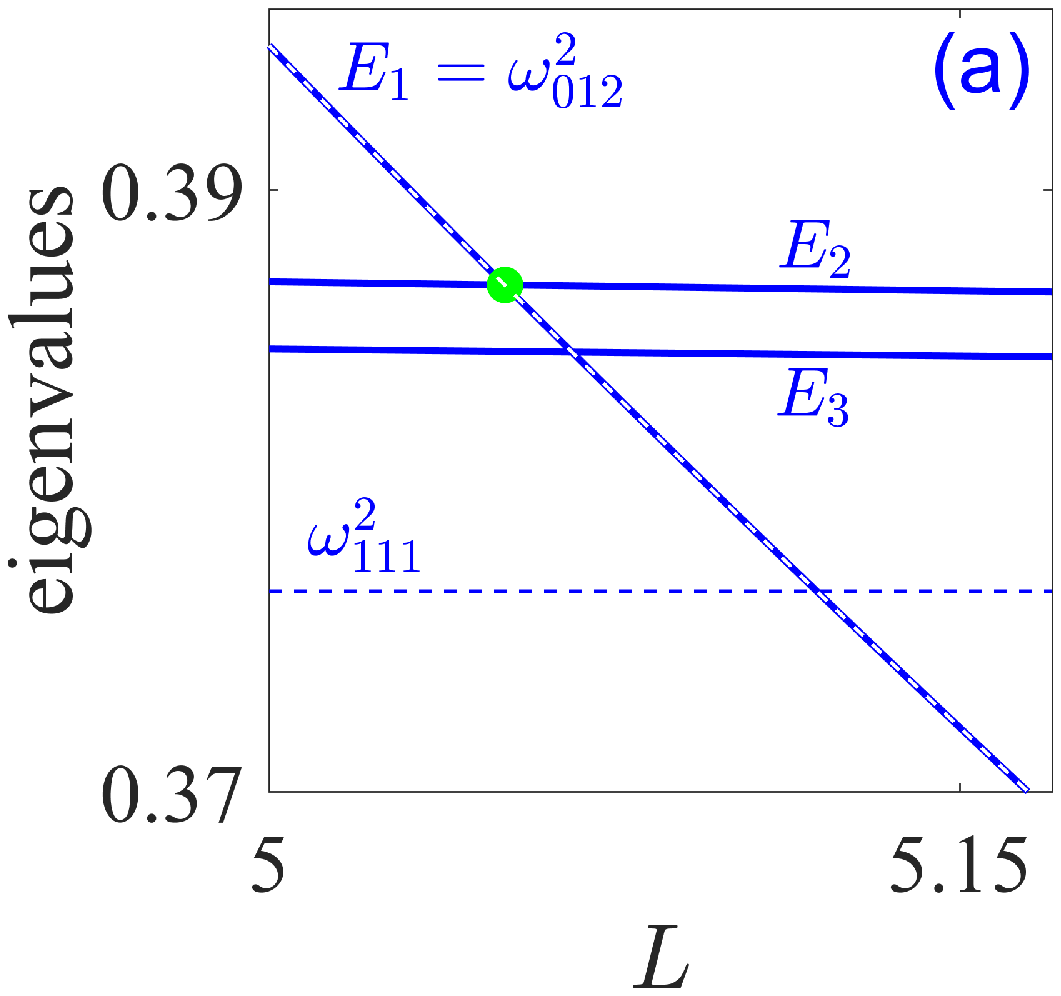}
\includegraphics[width=6cm,clip=]{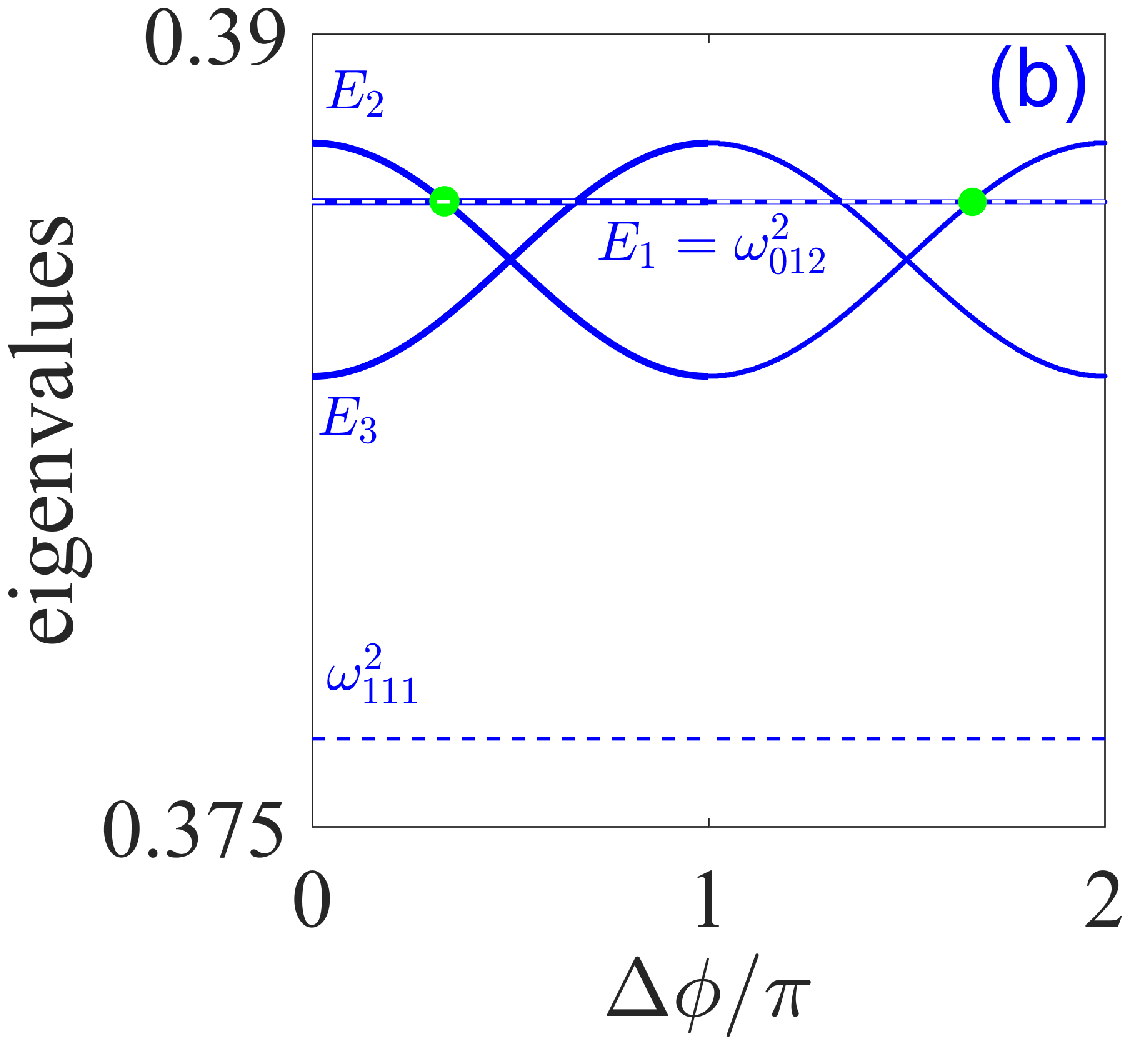}
\caption{(Color  online). The eigenvalues of the closed resonator (dash lines) and
the eigenlevels (\ref{neweig}) (solid lines) shifted by evanescent modes vs (a) the resonator length
at $\phi=\pi/3$ and (b) rotation angle at $L=5.0512$.}
\label{eigs}
\end{figure}

Let us consider the crossing of the eigenlevel
$\omega_{012}^2=\pi^2/L^2$ with the degenerate eigenlevel
$\omega_{111}^2=\mu_{11}^2/R^2$ shown in Fig. \ref{eigs} by dash
lines.
The coupling matrix elements of the eigenmodes with the first propagating channel
$p=0, q=1$ (see Table \ref{Tab1}) of the right waveguide
according to Eqs. (\ref{propag}), (\ref{eigRes}) and (\ref{Wp}) equal
\begin{eqnarray}\label{W01R}
&W_{mnl;01}=(w_0~ w_1~ w_1), w_0=W_{012;01}=\frac{1}{3}\sqrt{\frac{2}{L}},&\nonumber\\
&w_1=W_{\pm 111;01}=0.269\sqrt{\frac{1}{L}}&
\end{eqnarray}
for the given radius of the resonator. We also take into account
the coupling with the first evanescent modes $p=\pm 1, q=1$ of the
waveguide (see Table \ref{Tab1})
\begin{eqnarray}\label{W11R}
&W_{mnl;11}=(0~ v_1~ v_2), ~W_{mnl;-11}=(0~ v_2~ v_1),&\\
&v_1=W_{012;11}^L=0.1141\sqrt{\frac{1}{L}},
~v_2=W_{\pm 111;11}=-0.0141\sqrt{\frac{1}{L}}.&\nonumber
\end{eqnarray}
Because of the phase difference between the coupling matrix
elements for left and right waveguides we immediately obtain
\begin{eqnarray}\label{WL}
&W_{mnl;01}^R=(-w_0~ w_1e^{i\Delta\phi}~ w_1e^{-i\Delta\phi}),&\nonumber\\
&W_{mnl;11}^R=(0~ v_2e^{i\Delta\phi}~
v_1e^{-i\Delta\phi}),&\nonumber\\
&W_{mnl;-11}^R=(0~ v_1e^{i\Delta\phi}~v_2e^{-i\Delta\phi}).&
\end{eqnarray}
The contribution of the higher evanescent modes shown in Table
\ref{Tab1} is negligible. For open channel $p=0, q=1$ the wave
number $q_{01}=\omega$ while for the next closed channel $p=\pm 1,
q=1$ the wave number $k_{11}=iq_{11},
q_{11}=\sqrt{\mu_{11}^2-\omega^2}$ is imaginary. Then the
truncated effective Hamiltonian (\ref{Heff}) can be rewritten as
follows
\begin{equation}\label{Heff1}
\widehat{H}_{eff}=\widehat{H}_R+q_{11}\sum_{C=L,R} \sum_{p=\pm
1}\widehat{W}^C_{p=\pm 1,1}\{\widehat{W}^C_{p=\pm 1,1}\}^{\dagger}
-i\omega\sum_{C=L,R}\widehat{W}^C_{01}\{\widehat{W}^C_{01}\}^{\dagger}
=\widehat{\widetilde{H}}_R-i\omega\widehat{\Gamma},
\end{equation}
where the Hermitian term
\begin{equation}\label{HBSC1}
\widehat{\widetilde{H}}_R=\left(\begin{array}{ccc}
  \omega_{012}^2 & 0 & 0 \\
  0 & \omega_{111}^2+2q_{11}(v_1^2+v_2^2) & 2q_{11}v_1v_2(1+e^{-2i\Delta\phi}) \\
  0 & 2q_{11}v_1v_2(1+e^{2i\Delta\phi}) &  \omega_{111}^2+2q_{11}(v_1^2+v_2^2)\\
\end{array}\right)
\end{equation}
is the Hamiltonian of the resonator coupled to the evanescent
modes. The anti-Hermitian part takes the following form
\begin{equation}\label{GammaBSC1}
\widehat{\Gamma}=\left(\begin{array}{ccc}
  2w_0^2 & w_0w_1(1-e^{i\Delta\phi}) & w_0w_1(1-e^{-i\Delta\phi}) \\
  w_0w_1(1-e^{-i\Delta\phi}) & 2w_1^2 & w_1^2(1+e^{-2i\Delta\phi}) \\
  w_0w_1(1-e^{i\Delta\phi}) & w_1^2(1+e^{2i\Delta\phi}) &  2w_1^2\\
\end{array}\right).
\end{equation}
The eigenvalues of the Hamiltonian (\ref{HBSC1}) can be easily
found as
\begin{equation}\label{neweig}
E_1=\omega_{012}^2, E_{2,3}=\omega_{111}^2+2q_{11}[v_1^2+v_2^2\pm 2v_1v_2\cos\Delta\phi].
\end{equation}
Thus the evanescent modes of the waveguides non-coaxially attached
to the cylindrical resonator lift the degeneracy of eigenmodes
$\pm 1 1 1$ as shown in Fig. \ref{eigs} by solid lines. The only
case when the degeneracy is restored is the case
$\Delta\phi=\pi/2$.  The corresponding eigenmodes of the
Hamiltonian (\ref{HBSC1}) are the following
\begin{equation}\label{neweigmod}
 \mathbf{X}_1=   \left(\begin{array}{c}
  1 \\
  0 \\
  0 \\
\end{array}\right),
 \mathbf{X}_2=\frac{1}{\sqrt{2}}\left(\begin{array}{c}
  0 \\
  -e^{-i\Delta\phi} \\
  1 \\
\end{array}\right),
 \mathbf{X}_3=\frac{1}{\sqrt{2}}\left(\begin{array}{c}
  0 \\
  e^{-i\Delta\phi} \\
  1 \\
\end{array}\right).
\end{equation}

\begin{figure}[ht]
\includegraphics[width=6cm,clip=]{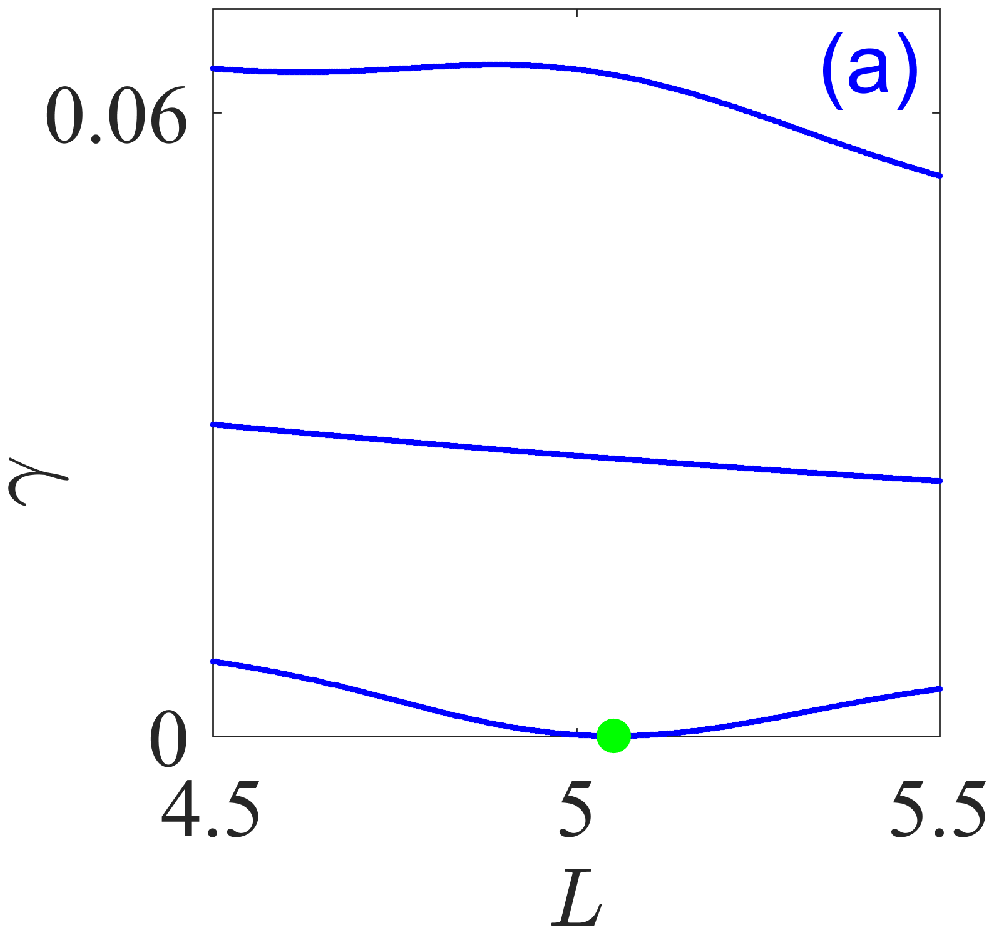}
\includegraphics[width=6cm,clip=]{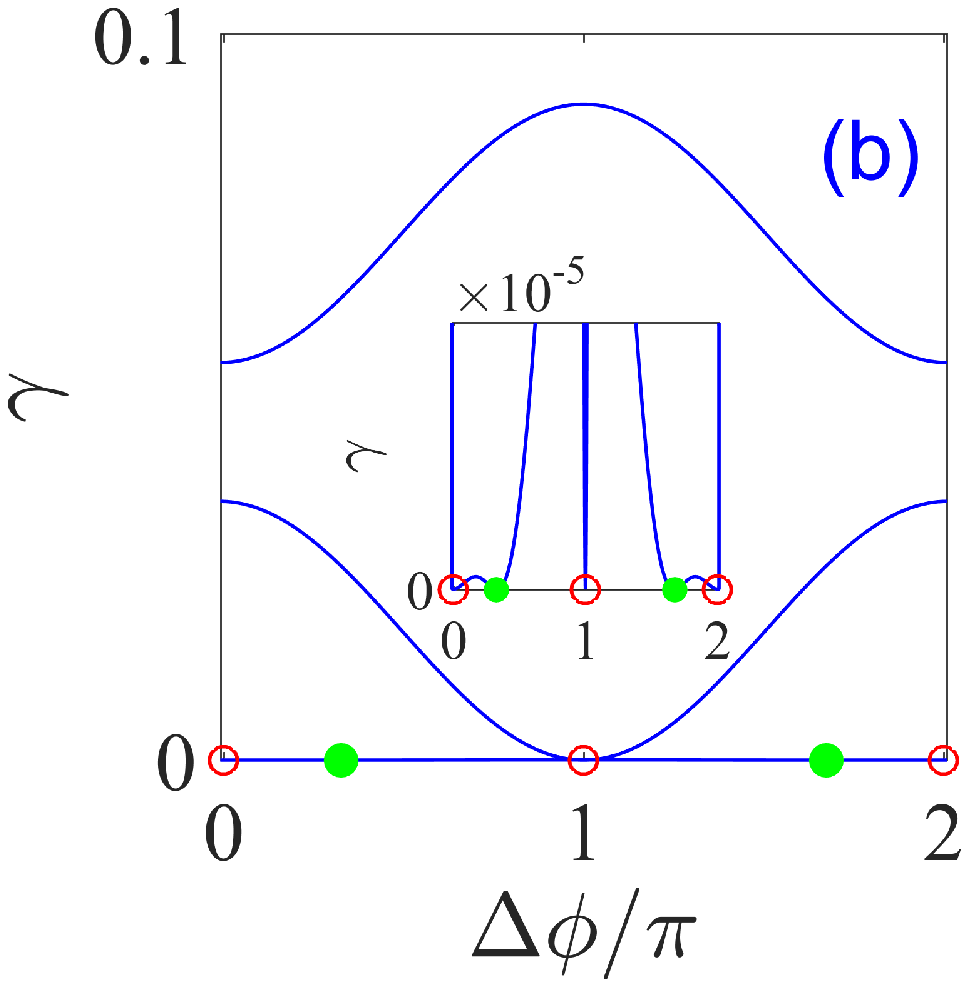}
\caption{(Color  online). The resonant width vs (a) the resonator length
at $\Delta\phi=\pi/3$ and (b) rotation angle at $L=5.0512$. Circles show the BSC
points.}
\label{gamma}
\end{figure}

Next, let us consider the BSC in the truncated version
(\ref{Heff1}). The point of the BSC can be easily diagnosed by
zero resonant width as shown in Fig. \ref{gamma}. For
$\Delta\phi=\pi/4$ the BSC occurs at $L=L_c=5.0512$ marked by
closed green circle in Fig. \ref{gamma} (a). Respectively at
$L=L_c$ the BSC occurs at $\Delta\phi=\pi/4$ and
$\Delta\phi=2\pi-\pi/4$. These points are seen in zoomed insert in
Fig. \ref{gamma} (b).

For $\Delta\phi=0$ both continua of left and right waveguides
coincide to result in the symmetry protected BSC superposed of
degenerate eigenmodes of the closed resonator $\psi_{111}$ and
$\psi_{-111}$ to be in the following form
\begin{equation}\label{sym prot BSC}
    \psi_{BSC}(r,\phi,z)=AJ_1(\mu_{11}r)\sin(\pi z/L)\sin\phi
\end{equation}
which always has zero coupling with the propagation mode
$\psi_{01}(\rho,\alpha,z)$ shown in Table \ref{Tab1}. As seen from
Eq. (\ref{sym prot BSC}) this conclusion also holds true for
$\Delta\phi=\pi$. This BSC is symmetry protected for arbitrary
resonator length as shown in Figs. \ref{gamma}, \ref{fig9}, and
\ref{fig10}.

As soon as $\Delta\phi\neq 0$ the continua become different to
destroy the symmetry protected BSCs. It could be expected that in
the case of two waveguides the point of threefold degeneracy where
the $\omega_{012}$ crosses the double degenerate $\omega_{111}$ as
shown in Fig. \ref{eigs} (a) is a BSC point in accordance with the
above consideration. However the BSC point where the resonant
width turns to zero (see Fig. \ref{gamma}) does not coincide with
this point. The computation on the basis of full basis effective
Hamiltonian gives the same result. In fact, the evanescent modes
split the eigenvalues (\ref{neweig}). Respectively the point of
threefold degeneracy $\omega_{111}^2=\omega_{012}^2(L)$ splits
into two double degenerate points $E_1(L)=E_2(L,\Delta\phi)$ and
$E_1(L)=E_3(L,\Delta\phi)$. As shown in Fig. \ref{eigs} (a) the
first case exactly corresponds to the BSC point but not the
second.

In the first case we can superpose the eigenmodes
(\ref{neweigmod}) as $a\mathbf{X}_1+b\mathbf{X}_2$ and require
zero coupling of this superposed mode with the left waveguide
\begin{equation}\label{coup1}
aw_0+\frac{b}{\sqrt{2}}w_1(1-e^{-i\Delta\phi})=0
\end{equation}
according to Eqs. (\ref{W01R}) and (\ref{neweigmod}). It is easy
to show that the coupling with the phase shifted continuum of the
left waveguide takes the {\it same} form as Eq. (\ref{coup1}).
Thus, the BSC has the following form
\begin{equation}\label{BSC1}
\psi_{BSC}=w_1(1-e^{-i\Delta\phi})\psi_{012}+w_0(e^{-i\Delta\phi}\psi_{111}-\psi_{-111}).
\end{equation}
Substituting eigenmodes (\ref{eigRes}) we obtain
\begin{equation}\label{BSC2}
\psi_{BSC}=2ie^{-i\Delta\phi/2}[w_1\sin(\Delta\phi/2)\psi_{01}(r)\psi_2(z)+
w_0\sin(\phi-\Delta\phi/2)\psi_{11}(r)\psi_1(z)].
\end{equation}
One can see that this BSC does not support spinning currents of
acoustic intensity $\overrightarrow{j}=\psi^{*}\nabla\psi$ in
contrast to coaxial waveguides \cite{Duan}. This is due to
evanescent modes of the non-coaxial waveguides which lift the
degeneracy of the eigenmodes $\Psi_{mnl}$ with respect to the sign
$m$.

\begin{figure}[ht]
\includegraphics[width=10cm,clip=]{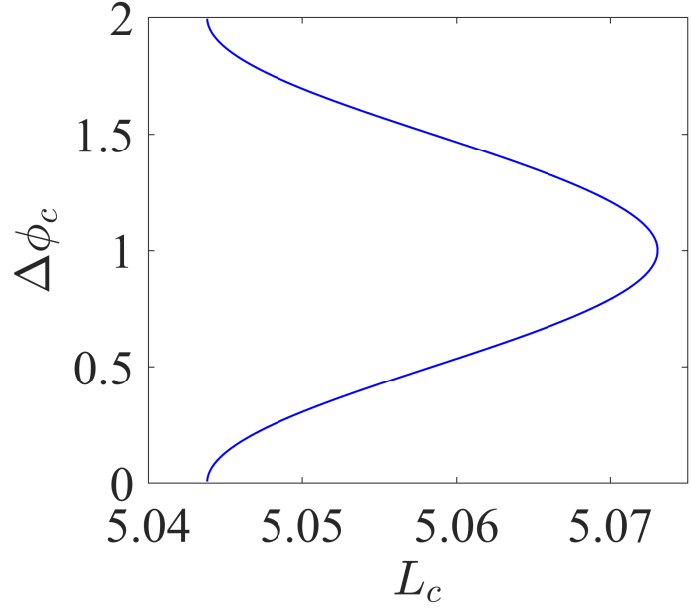}
\caption{(Color  online) Line of the BSCs in the parametric space of the
resonator length and rotation angle $\Delta\phi$.}
\label{fig8}
\end{figure}

The BSC point is given by the equation $E_1(L)=E_2(L,\Delta\phi)$
which gives rise to a line of the BSC in the parametric space $L$
and $\Delta\phi$ shown in Fig. \ref{fig8}.
Thus, the only phase difference between the continua allow the
BSCs in the point of twofold degeneracy. This is necessary for
existence of BSC but not sufficient. Indeed let us consider the
another point of degeneracy $E_1=E_3$ (see Fig. \ref{eigs} (a)).
At this point we adjust the superposition
$a\mathbf{X}_1+b\mathbf{X}_3$ for cancellation of the coupling
with both continua. The analogue of Eq. (\ref{coup1}) takes the
following form
\begin{equation}\label{coup2}
\pm aw_0+\frac{b}{\sqrt{2}}w_1(1+e^{i\Delta\phi})w_1=0.
\end{equation}
These equations can not be fulfilled simultaneously to forbid
this degeneracy point as the BSC point.

\begin{figure}[ht]
\includegraphics[width=12cm,clip=]{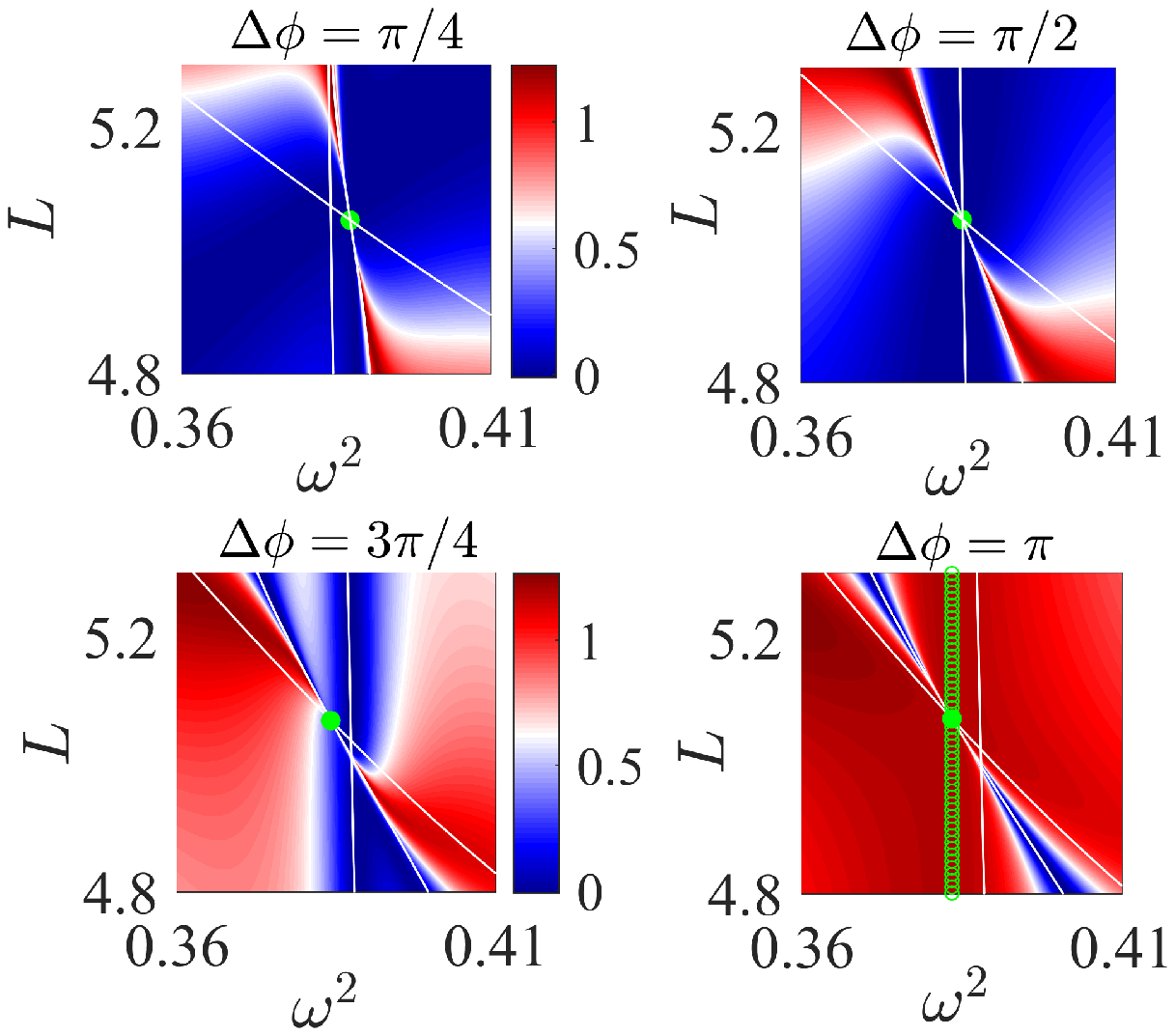}
\caption{(Color  online) (a)
Transmittance vs frequency and resonator length at four fixed rotation angles.
Solid green lines show the resonances defined by real part of the complex eigenvalues of
the effective Hamiltonian (\ref{Heff1}). Closed circles mark the BSCs which exactly
correspond to points of degeneracy of the eigenlevels (\ref{neweig}).}
\label{fig6}
\end{figure}

\begin{figure}[ht]
\includegraphics[width=12cm,clip=]{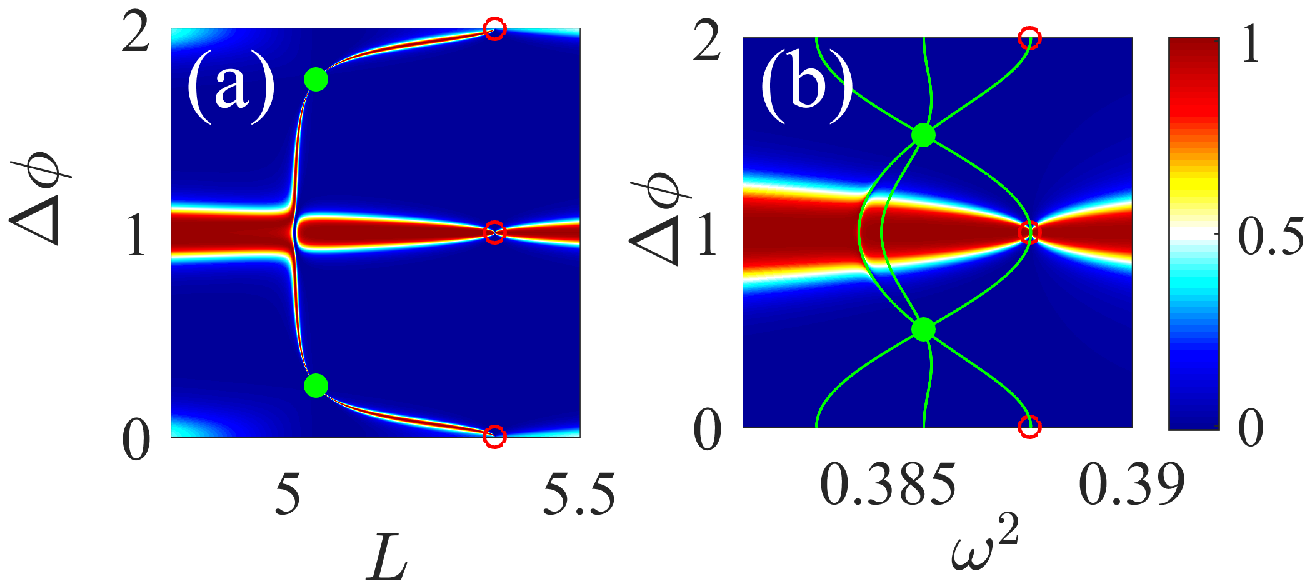}
\caption{(Color  online) (a) Transmittance vs the resonator
length and rotation angle for the frequency tuned onto the
frequency of the BSC  $\omega_c^2=0.388$. (b) Transmittance vs the
frequency and rotation angle for the length tuned onto the BSC
length $L_c=5.048$. Closed circles mark BSC 1 resulted by crossing
of eigenlevels (\ref{neweig}) $E_1$ and $E_2$, open circles mark
the symmetry protected BSCs (\ref{sym prot BSC}).}
\label{fig9}
\end{figure}

\begin{figure}[ht]
\includegraphics[width=12cm,clip=]{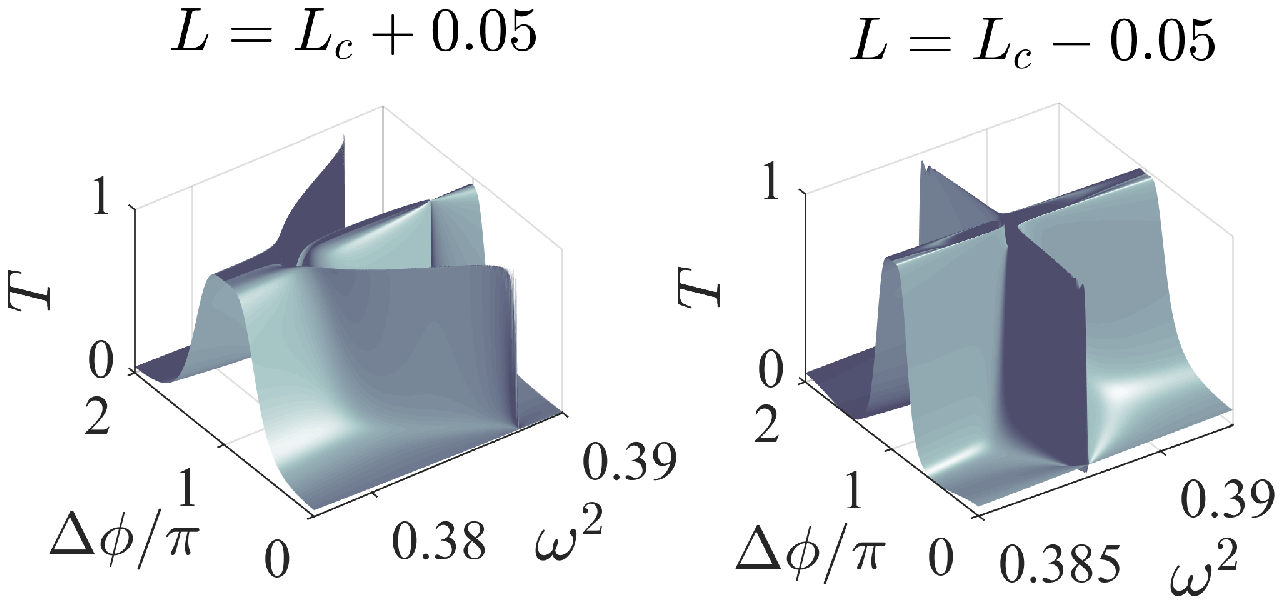}
\caption{(Color  online)
Transmittance vs the frequency and rotation angle for
the length detuned from the BSC length $L_c$.}
\label{fig10}
\end{figure}

By the use of Eq. (\ref{S-matrix}) and truncated effective
Hamiltonian (\ref{Heff1}) we calculated the transmittance with the
results presented in Fig. \ref{fig6}. Comparison to Fig.
\ref{fig2} (a) and (b) shows that all features of the
transmittance can be well reproduced in the vicinity of the BSCs
by the use of truncated basis.
One can also see from Fig. \ref{fig6} that the resonant features
follow the real parts of the complex eigenvalues of the effective
non-Hermitian Hamiltonian (\ref{Heff1}) when $\Delta\phi\neq 0$.

Fig. \ref{fig9} shows fine features of the transmittance vs two
parameters for the third parameter exactly tuned to the BSC. Fig.
\ref{fig9} (a) demonstrates a Fano resonance collapse in the
parametric space of length and rotation angle at the BSC point
$L_c=5.048$ and $\Delta\phi_c=\pi/4$ with the frequency exactly
tuned to the BSC $\omega_c=0.3873$. Fig. \ref{fig9} (b) shows the
transmittance vs the frequency and the rotation angle for the
length of the resonator tuned to the BSC length $L_c=5.0584$. Fig.
\ref{fig9} (a) and Fig. \ref{fig9} (b) show that the resonator is
blocked when $\Delta\phi=0$ and open when $\Delta\phi=\pi$. Fig.
\ref{fig10} demonstrates as the transmittance is sensitive to
small deviations from the BSC length $L_c$.

\begin{figure}[ht]
\includegraphics[width=12cm,clip=]{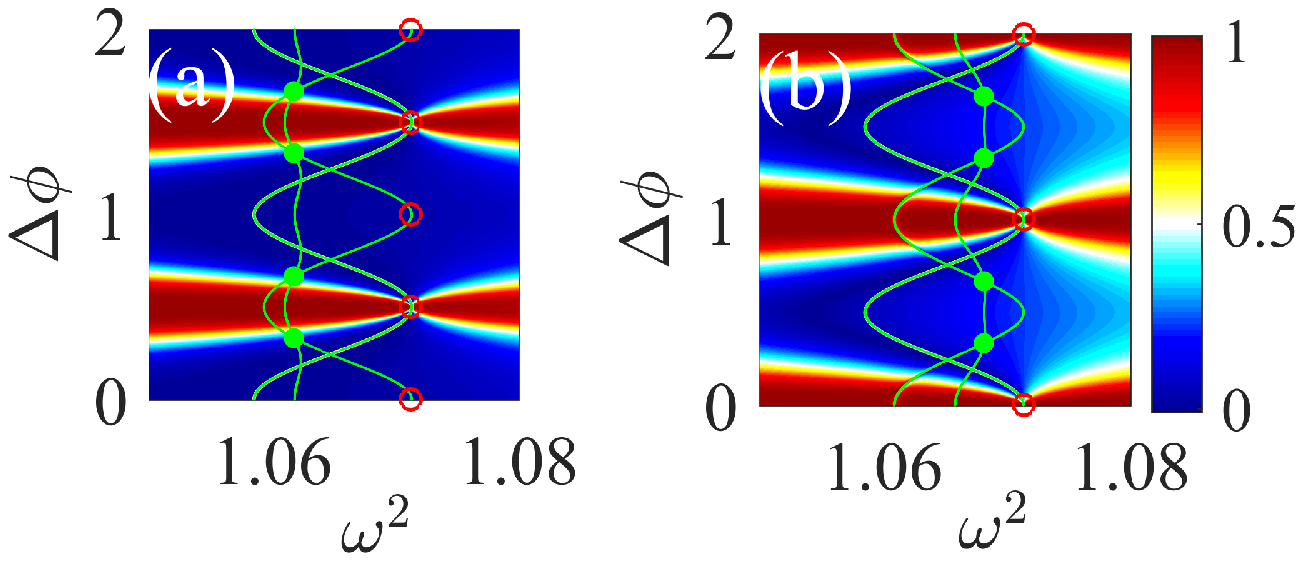}
\caption{(Color  online)
Transmittance vs the frequency and rotation angle in the vicinity of
crossing of the modes (a) $012$ and $211$ at $L=3$ and (b) $013$ and $211$, $L=4$.}
\label{fig11}
\end{figure}

One can see from Eqs. (\ref{W01R}) and (\ref{WL}) that at
$\Delta\phi=0$ the channels $012$ and $\pm 1 11$ interfere
destructively at the right output to block the wave transmission
through the resonator. In contrast for $\Delta\phi=\pi$ the
channels interfere constructively at the right output to maximize
the transmittance. Along the same line for the channels $012$ and
$\pm 2 11$ the wave faucet will open at $\Delta\phi=\pi/2, 3\pi/2$
because of the phase factor $e^{\pm 2i\Delta\phi}$ in the coupling
matrix elements (\ref{WL}). Respectively, interference of channels
of $013$ and $\pm 2 11$ will realize the wave faucet which opens
at $\Delta\phi=0,\pi$. Fig. \ref{fig11} completely confirms the
above predictions.
\subsection{The mode $\pm 112$ crosses the mode $\pm 211$}
The coupling matrix elements of the eigenmodes with the first
propagating channel $p=0, q=1$ (see Table \ref{Tab1}) of the right
waveguide according to Eqs. (\ref{propag}), (\ref{eigRes}), and
(\ref{Wp}) equal
\begin{eqnarray}\label{WW01R}
&W_{mnl;01}^R=(w_1~ w_1~ w_2 ~w_2), w_1=W_{211;01}^R=0.1737\sqrt{\frac{1}{L}},&\nonumber\\
&w_2=W_{\pm 112;01}^R=0.2849\sqrt{\frac{2}{L}}.&
\end{eqnarray}
The coupling matrix elements with the first evanescent modes $p=1,
n=1$ of the left waveguide (see Table \ref{Tab1}) equal
\begin{eqnarray}\label{WW11R}
&W_{mnl;-11}^L=(v_1~ v_2~ v_3~ v_4),~W_{mnl;11}^L=(v_2~ v_1~ v_4~ v_3), ,&\\
&v_1=W_{-211;11}^L=0.2197\sqrt{\frac{1}{L}}, ~v_2=W_{211;11}^L=-0.0187\sqrt{\frac{1}{L}},&\nonumber\\
&v_3=W_{-112;11}^L=0.1709\sqrt{\frac{2}{L}},
~v_4=W_{112;11}^R=-0.0157\sqrt{\frac{2}{L}}.&\nonumber
\end{eqnarray}
Respectively according to Eq. (\ref{WLWR}) we have for the phase
shifted coupling matrix elements
$W_{mnl;pq}^R=V_{mnl}W_{mnl;pq}^L$. The resulting effective
non-Hermitian Hamiltonian (\ref{Heff})
$$\widehat{H}_{eff}=\widehat{H}-i\omega\widehat{\Gamma}$$ consists
of
\begin{equation}\label{HBSC34}
\widehat{\widetilde{H}}=q_{11}\left(\begin{array}{cccc}
  \omega_{211}^2/q_{11}+2v_1^2 & v_1v_2(1+e^{4i\Delta\phi}) &v_1v_3(1-e^{-i\Delta\phi})
  & v_1v_4(1-e^{-3i\Delta\phi}) \\
  v_1v_2(1+e^{-4i\Delta\phi}) & \omega_{211}^2/q_{11}+2v_2^2 & v_2v_3(1-e^{3i\Delta\phi}) &
  v_1v_4(1-e^{i\Delta\phi}) \\
  v_1v_3(1-e^{i\Delta\phi}) & v_2v_3(1-e^{-3i\Delta\phi}) & \omega_{112}^2/q_{11}+2v_3^2 &
   v_3v_4(1+e^{-2i\Delta\phi})\\
v_1v_4(1-e^{3i\Delta\phi}) & v_2v_4(1-e^{-4i\Delta\phi}) & v_3v_4(1+e^{2i\Delta\phi}) &
\omega_{112}^2/q_{11}+2v_4^2\\
\end{array}\right)
\end{equation}
and
\begin{equation}\label{GammaBSC34}
\widehat{\Gamma}=\left(\begin{array}{cccc}
  2|w_1|^2 & w_1^2(1+e^{4i\Delta\phi})&  w_1w_2(1-e^{-i\Delta\phi})   & w_1w_2(1-e^{-3i\Delta\phi})\\
  w_1^2(1+e^{-4i\Delta\phi}) & 2w_1^2 & w_1w_2(1-e^{3i\Delta\phi})   & w_1w_2(1-e^{i\Delta\phi})\\
    w_1w_2(1-e^{i\Delta\phi})  & w_1w_2(1-e^{-3i\Delta\phi}) &  2w_2^2 & w_2^2(1+e^{-2i\Delta\phi})\\
w_1w_2(1-e^{3i\Delta\phi}) &w_1w_2(1-e^{-i\Delta\phi}) & w_2^2(1+e^{2i\Delta\phi})& 2w_2^2
\end{array}\right).
\end{equation}

\begin{figure}[ht]
\includegraphics[width=10cm,clip=]{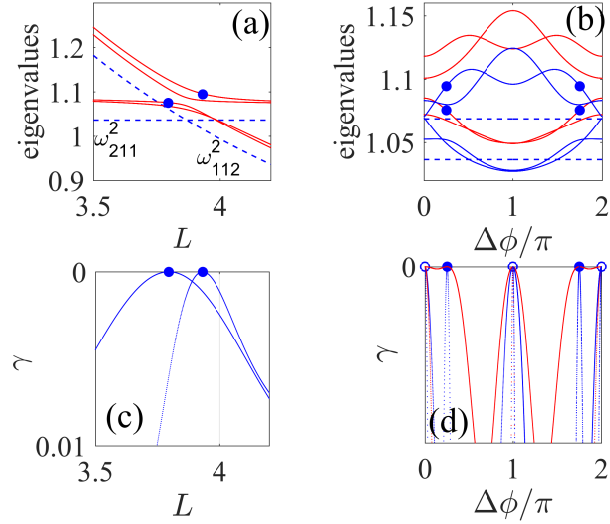}
\caption{(Color  online) The eigenlevels (\ref{eig34}) (solid lines) as dependent on (a) length of resonator
at $\phi=\pi/41$ and (b) rotation angle at $L=3.7947$ (blue) and $L=3.9312$ (green) compared to the eigenfrequencies of the
closed resonator (dash lines). (c) and (d) Corresponding behavior of the resonant
widths defined by imaginary parts of the effective non-Hermitian
Hamiltonian (\ref{HBSC34}) and (\ref{GammaBSC34}).
Closed circles mark the BSCs. Open circles mark the symmetry protected BSCs.}
\label{eig34}
\end{figure}

In contrast to the previous case of crossing of eigenlevels $112$
and $012$ (the BSC 1), the present case of full matrices impedes
analytical consideration of the BSCs. Nevertheless the small size
of matrices (\ref{HBSC34}) and (\ref{GammaBSC34}) facilitates
numerical treatment of  the BSCs. Fig. \ref{eig34} shows the
eigenlevels of the Hamiltonian (\ref{HBSC34})
\begin{equation}\label{Heig}
\widehat{\widetilde{H}}\mathbf{X}_j=E_j\mathbf{X}_j
\end{equation}
as dependent on the length and rotation angle. One can see that
the presence of evanescent modes lifts the  degeneracy relative to
the sign of $m$.
Below in Fig. \ref{eig34} we show the resonant widths that
demonstrates two BSCs at the points: 1) $L_c=3.7947,
\omega_c^2=1.0756, \phi_c=\pi/4$ and 2) $L_c=3.9312,
\omega_c^2=1.0946, \phi=\pi/4$. Respectively, in Fig. \ref{eig34}
(b) we show the eigenvalues for these BSC lengths.

While the former BSCs occur at the crossing
of eigenlevels (\ref{neweig}) the present BSCs are allocated
neither at the crossing of neither the eigenfrequencies
$\omega_{112}$ and $\omega_{211}$ nor the eigenlevels modified by
the evanescent modes as seen from Fig. \ref{eig34}. Such
phenomenon is generic in open chaotic billiards where the
eigenlevels of the closed billiard undergo avoided crossing
\cite{Pilipchuk} similar to that shown in Fig. \ref{eig34} (a). In
that case some of the eigenmodes of the Hamiltonian (\ref{HBSC34})
can decouple under the evolution of the parameters of the
resonator. In the truncated description of the BSCs we obtain the
following solutions for the BSCs presented in Table \ref{Tab3}.
\begin{table}
\caption{BSC solutions  3 and 4 resulted by crossing of modes $\pm 112$ and $\pm 211$.}
\begin{tabular}{|c|c|c|c|c|c|}
  \hline
BSC number & $\omega^2$ & $L$ & $\Delta\phi$ & $mnl$& $a_{mnl}$ \\
  \hline
 3& 1.0756& 3.7947 & $\pi/4$ & 211 &  0.6715i \\
&       &       &         &-211 &0.6715\\
&       &       &         &112 &-0.2047-0.0848i\\
 &       &       &         &-112 &-0.0848-0.2047i\\
\hline
 &       &       &         & 211 & -0.553i\\
4& 1.0966& 3.9312 & $\pi/4$ & -211 &  0.553 \\
 &       &       &         &112 &-0.407-0.168i\\
 &       &       &         &-112 &0.168+0.407i\\
\hline
 \end{tabular}
\label{Tab3}
\end{table}
Comparison with numerical results in full basis (Table \ref{Tab2})
shows good agreement with the BSCs (2) and (3). Anyway as seen
from Fig. \ref{fig5} the BSCs are expanded over all four
eigenmodes $\pm 211$ and $\pm 112$ of the closed resonator.
Remarkably, in the description of the eigenmodes (\ref{Heig})
$\mathbf{X}_j$ we obtain that the numerical amplitudes of the BSC
mode $\psi_{BSC}=\sum_{j=1}^4b_j\mathbf{X}_j$ are the following
\begin{figure}[ht]
\includegraphics[width=6cm,clip=]{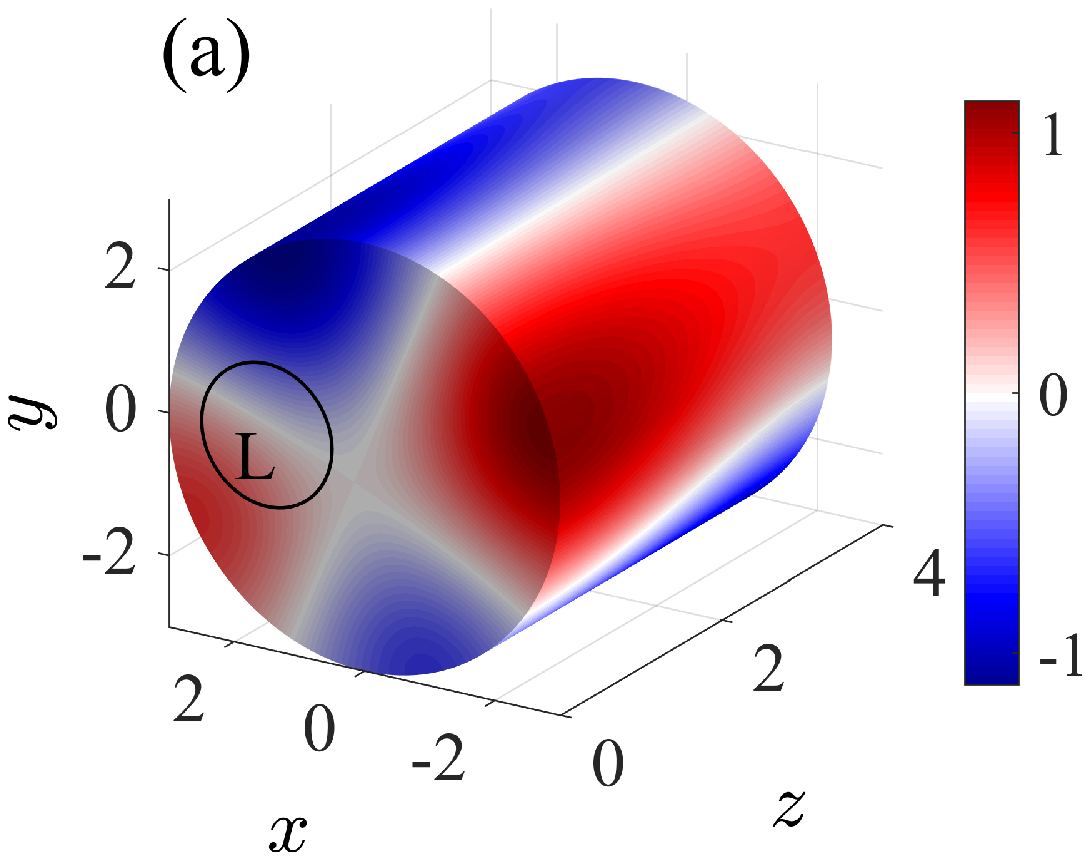}
\includegraphics[width=6cm,clip=]{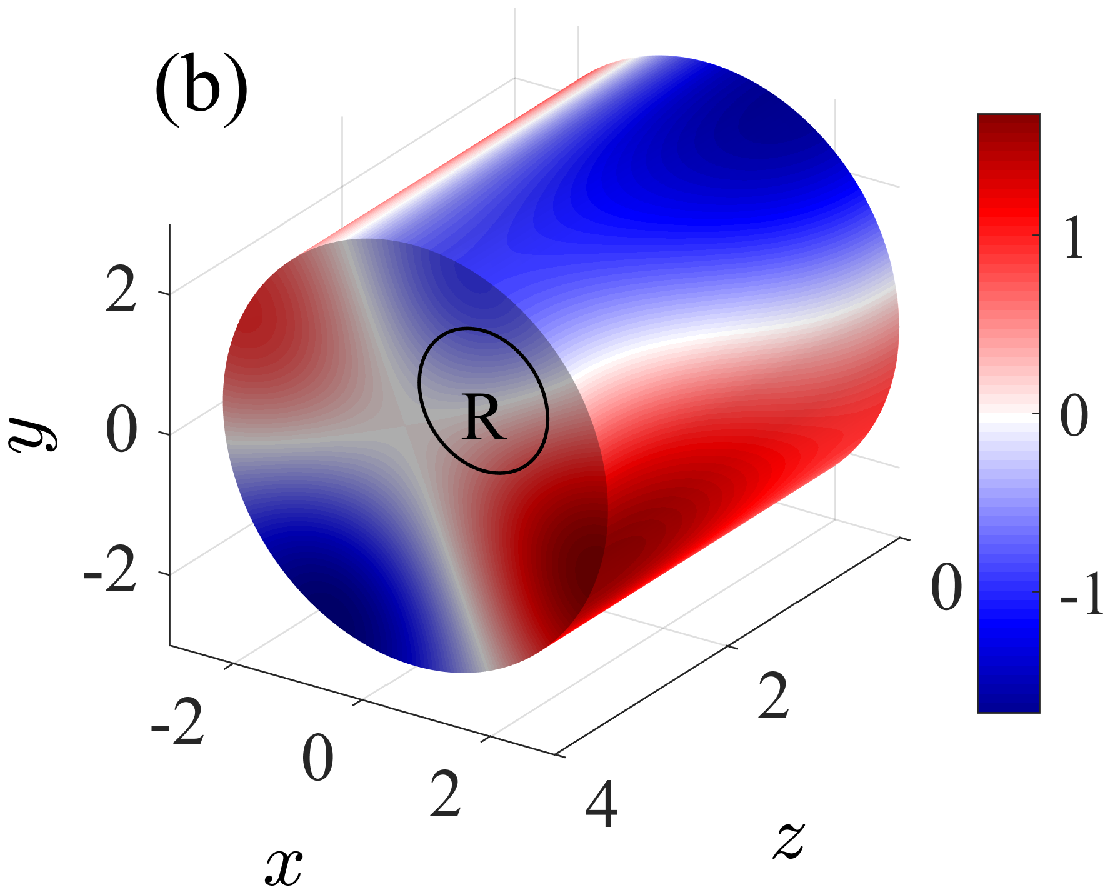}
\includegraphics[width=6cm,clip=]{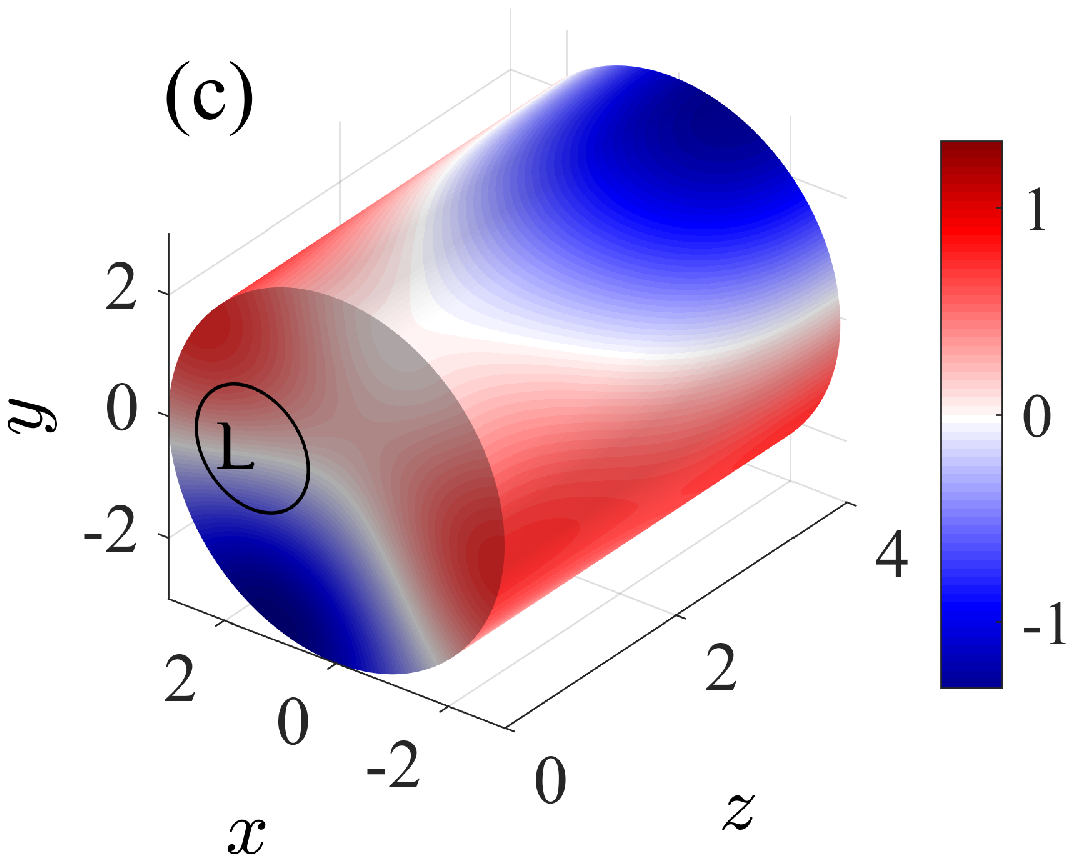}
\includegraphics[width=6cm,clip=]{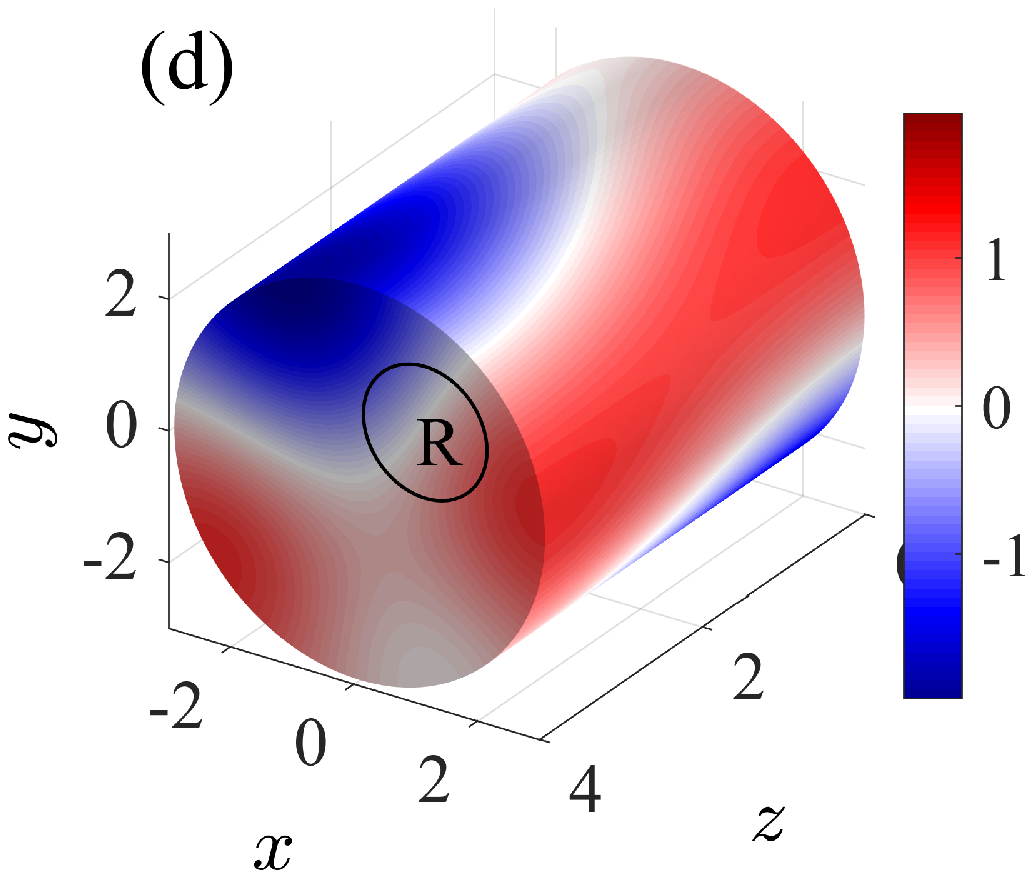}
\caption{(Color  online) The real parts of the eigenmodes (\ref{Heig}) of modified resonator.
(a) and (b) $\mathbf{X}_2$
at the BSC point (3) (Eq. (\ref{BSCfourX})) with side of input
and output respectively. (c) and (d) the same for the $\mathbf{X}_1$ at
the BSC point (4).}
\label{BSC34}
\end{figure}
\begin{eqnarray}\label{BSCfourX}
    &(3)~~ \psi_{BSC}=\mathbf{X}_2, ~L_c=3.7947, ~\omega_c^2=1.0756, ~\Delta\phi_c=\pi/4,&\nonumber\\
    &(4)~~ \psi_{BSC}=\mathbf{X}_1, ~L_c=3.9312, ~\omega_c^2=1.0966, ~\Delta\phi_c=\pi/4.&
\end{eqnarray}
Thus, in this description we have an important result that the BSC
is simply one of the eigenmodes of the resonator coupled to the
evanescent modes. For this eigenmode to be decoupled from the
propagating channel of both waveguides  the overlapping integrals
\begin{equation}\label{WpX}
W^C_{j;01}=\psi_{01}(z=z_C)\psi_l(z_C)\int_0^{2\pi} d\alpha\int_0^1\rho d\rho
\overrightarrow{X}_j(r(\rho,\alpha),\phi(\rho,\alpha)), C=L,R
\end{equation}
have to be zero for selected $j=1$ or $j=2$.
Fig. \ref{BSC34} demonstrates that the overlapping integral
vanishes at the BSC points. Such BSCs form new class of the
accidental BSCs \cite{Pilipchuk,anneal,Bo Zhen}. Fig. \ref{BSC34}
demonstrates that the BSC modes $\overrightarrow{X}_j$ are twisted
to have zero coupling on both interfaces.
\begin{figure}[ht]
\includegraphics[width=8cm,clip=]{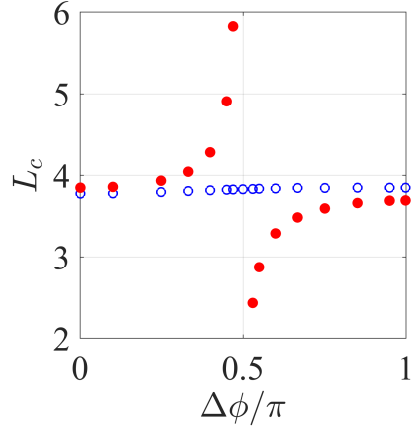}
\caption{(Color  online) Lines of the BSCs in the parametric space of the
resonator length and rotation angle $\Delta\phi$.}
\label{Lcphi}
\end{figure}
Again the BSCs occur on lines in the parametric space $L$ and
$\Delta\phi$ as shown in Fig. \ref{Lcphi} by open circles. However
there is the second line in which the BSC length $L_c$ diverges
when $\Delta\phi\rightarrow \pi/2$. Also it is important to note
that the BSC modes do not support currents of
acoustic intensity similar to the BSC 1 and 2.
\section{Conclusions}
The method of the effective non-Hermitian Hamiltonian equivalent to the coupled mode
theory \cite{Maksimov} calculates the transmittance of waves through the resonator by
Eq. (\ref{S-matrix}). One can see that the shape of resonances and interference of them
depends not only on  the eigenfrequencies but also on the coupling matrix elements
of the corresponding eigenmodes with propagating modes of the waveguides attached to the
resonator. That effect was first experimentally demonstrated by Rotter {\it et al}
\cite{Kuhl} by change of the coupling strength with use of diaphragms between the waveguides
and the resonator. In the present paper we propose the path through which the coupling matrix
elements do not change the strength but acquire only phase factors by simple means of
rotation of input waveguide relative to the resonator.
Other words, that transforms phase of
the continua of the waveguide and can be performed in a realistic acoustic or electromagnetic
experiment by the use of piston-like hollow-stem waveguides tightly fit
to the interior boundaries of a cylindric cavity \cite{Lyapina} as shown in Fig. \ref{fig1}.
Our numerical and analytical calculations have shown that simple setup cardinally effect
Fano resonances that in turn opens and closes wave transmittance by small rotation
to be claimed as the wave faucet.

Rigorously speaking rotation of one waveguides in a way shown in Fig. \ref{fig1}
the number of continua is doubled at $\Delta\phi\neq 0$.
The problem of the BSC residing in a finite number of continua was considered
in Ref. \cite{Pavlov} in the framework of the Weisskopf-Wigner model \cite{Feshbach}.
Rigorous statement about the BSCs was formulated as follows.
The interference among $N$ degenerate states which decay into K non-interacting continua
generally leads to the formation of $N-K$ BSCs.
The equivalent point of view is that the linear superposition of the $N$
degenerate eigenmodes can be achieved to
have zero coupling with $K$ different continua  in $N-K$ ways by a variation of the
$N$ superposition coefficients $a_n$ \cite{JPhys28}.
In particular if waveguides were attached at different radiuses the BSC would need in triple degeneracy
of eigenlevels. However in the present case of rotation of one of the waveguides
attributes only the phase factor $\exp(im\Delta\phi)$ to the coupling matrix elements
 of resonator with the left continua  as was derived in Eq. (\ref{WLWR}).
That allowed to avoid problem of triple degeneracy of resonator eigemodes.
Moreover it gave rise to surprisingly rich variety of BSCs and respectively
collapses of Fano resonances.

Off-center attachment of waveguides lifts a degeneracy of eigenmodes of the closed cylindrical
resonator relative to sign of OAM given by integer $m$. As result the BSC is realized
in the vicinity degeneracy of eigenlevels modified by evanescent modes with definite $m$ and mode
$m=0$. Respectively the BSC expanded over the eigenmodes of closed resonator holds different contributions
of $m$ and $-m$ as explicitly  given for example in Eq. (\ref{BSC1}).

To be specific we consider the case of acoustic wave transmission
with the Neumann BCs but similar consideration can be performed
for electromagnetic wave transmission with the Dirichlet BCs.

{\bf Acknowledgement}: This work has been supported by RFBR through Grant 17-02-00440.
A.S. acknowledges H. Schanz, P. Seba, L. Sirko and H.-J. St\"{o}ckmann for encouraging
discussions.
The authors thank D.N. Maksimov for helpful discussions.


\begin{thebibliography}{99}
\bibitem{Young}
Thomas Young, "Bakerian Lecture: Experiments and calculations relative to physical optics",
Phil. Trans. Royal Soc.\textbf{94}, 1–16 (1804). doi:10.1098/rstl.1804.0001.
\bibitem{Neumann}
J. Von Neumann and E. Wigner, "\"{U}ber merkw\"{u}rdige diskrete eigenwerte,"
Z. Phys \textbf{30}, 467 (1929).
\bibitem{ABO}
Y. Aharonov and D. Bohm, "Significance of Electromagnetic Potentials in the Quantum
Theory", Phys. Rev. \textbf{115}, 485 (1959).
\bibitem{Fano}
U. Fano, "Effects of Configuration interactions on Intensities and Phase Shifts",
Phys. Rev. \textbf{124}, 1866 (1961).
\bibitem{Nye}
J.F. Nye and M.V. Berry, "Dislocations in wave trains",
Proc. R. Soc. London, Ser.A \textbf{336}, 165 (1974).
\bibitem{Lyapina}
A.A. Lyapina, D.N. Maksimov, A.S. Pilipchuk, and A.F. Sadreev, "Bound
states in the continuum in open acoustic resonators," J.Fluid Mech. \textbf{780}, 370 (2015).
\bibitem{Kuhl}
S. Rotter, F. Libisch, J. Burgoffer, U. Kuhl, and H.-J. St\"{o}ckmann, "Tunable Fano
resonances in transport through microwave billiards,"
Phys. Rev. E\textbf{69}, 046208 (2004).
\bibitem{Friedrich}
H. Friedrich and D. Wintgen, "Interferring resonances and bound states in the continuum,"
Phys. Rev. A{\bf 32}, 3231 (1985).
\bibitem{Volya}
A. Volya and V. Zelevinsky, "Non-Hermitian effective Hamiltonian and continuum shell model,"
Phys. Rev. C\textbf{67}, 054322 (2003).
\bibitem{SBR}
A.F. Sadreev, E.N. Bulgakov, and I. Rotter, "Bound states in the continuum in open quantum
billiards with a variable shape," Phys. Rev. B {\bf 73}, 235342 (2006).
\bibitem{Parker}
R. Parker, "Acoustic resonances and blade vibration in axial flow compressors",
J. Sound  and  Vibration \textbf{92}, 529 (1984).
\bibitem{Duan}
Y. Duan and M. McIver, "Rotational  acoustic resonances in cylindrical waveguides",
Wave Motion \textbf{39}, 261 (2004).
\bibitem{Hsu review}
Chia Wei Hsu, Bo Zhen, A.D. Stone, J.D. Joannopoulos and M. Solja\v{c}i\'{c},
"Bound states in the continuum," Nature Rev. Mat. \textbf{1} (2016).
\bibitem{Kante}
A. Kodigala, T. Lepetit, Q. Gu, B. Bahari, Y. Fainman, and B. Kant\'{e}, "Lasing action from
photonic bound states in continuum", Nature \textbf{541}, 196--199 (2017).
\bibitem{Zhang}
ZH. Zhang, Y. Li, W. Liu, J. Yang, Y. Ma, H. Lu,  Y. Sun,  H. Jiang, "Controllable lasing behavior enabled by
compound dielectric waveguide grating structures", Opt. Express \textbf{24}, 19458--19466 (2016).
\bibitem{Yoon}
J.W. Yoon, S.H. Song, R. Magnusson, "Critical field enhancement of asymptotic optical bound states in the
continuum,"  Sci. Rep. \textbf{5}, 18301 (2015).
\bibitem{Zhang&Zhang}
Mingda Zhang and Xiangdong Zhang, "Ultrasensitive optical absorption in
graphene based on bound states in the continuum," Sci. Rep. \textbf{5}, 8266, (2015).
\bibitem{Appl_Sc}
E.N. Bulgakov, A.F. Sadreev, and D.N. Maksimov, "Light Trapping
above the Light Cone in One-Dimensional Arrays of Dielectric
Spheres", Appl. Sciences \textbf{7}, 147 (2017).
\bibitem{Arxiv}
E.N. Bulgakov, and D.N. Maksimov, "Light enhancement by dielectric
arrays," arXiv preprint arXiv:1702.05990 (2017).
\bibitem{Rotter}
J. Oko{\l}owicz, M. P{\l}oszajczak, and I. Rotter, "Dynamics of quantum systems embedded
in a continuum," Phys. Rep. {\bf 374}, 271--383 (2003).
\bibitem{SR}
A.F. Sadreev and I. Rotter, "S-matrix theory for transmission through billiards in
tight-binding approach," J. Phys. A {\bf 36}, 11413 (2003).
\bibitem{Maksimov}
D. N. Maksimov, A. F. Sadreev, A. A. Lyapina, and A. S. Pilipchuk,
"Coupled mode theory for acoustic resonators," Wave Motion, \textbf{56}, 52 (2015).
\bibitem{Pichugin}
K. Pichugin, H. Schanz, and P. Seba, "Effective coupling for open billiards,"
Phys. Rev. E\textbf{64}, 056227 (2001).
\bibitem{RS2005}
I. Rotter and A. F. Sadreev, "Zeros in single-channel transmission through double quantum dots,"
Phys. Rev. E \textbf{71}, 046204 (2005).
\bibitem{Sok&Zel}
V.V. Sokolov and V.G. Zelevinsky, "Dynamics and statistics of unstable quantum states," Nucl. Phys. A \textbf{504}, 562--588 (1989).
\bibitem{Lepetit2014}
T. Lepetit and B. Kant'{e}, "Controlling multipolar radiation with symmetries
for electromagnetic bound states in the continuum," Phys. Rev. B\textbf{90},  241103(R) (2014).
\bibitem{Cobelli}
P. J.  Cobelli, V. Pagneux, A. Maurel, and P. Petitjeans, "Experimental study on water-wave trapped
modes," J. Fluid Mech. \textbf{666}, 445 (2011).
\bibitem{Pavlov}
F. Remacle, M. Munster, V.B. Pavlov-Verevkin and M. Desouter-Lecomte,
"Trapping in competitive decay of degenerate states," Phys. Lett. A \textbf{145}, 265 (1990).
\bibitem{Feshbach}
H. Feshbach, "Unified Theory of Nuclear reactions," Ann. Phys. N.Y. {\bf 5} 357 (1958).
\bibitem{JPC28}
E.N. Bulgakov and A.F. Sadreev, "Spin polarized bound states in the continuum in open Aharonov–Bohm
rings with the Rashba spin–orbit interaction," J. Phys.: Cond. Mat. \textbf{28}, 265301 (2016).
\bibitem{anneal}
E. Bulgakov and A. Sadreev, "Formation of bound states in the continuum for a quantum dot with
variable width," Phys. Rev. B \textbf{83}, 235321 (2011).
\bibitem{Wei}
Chia Wei Hsu, Bo Zhen, J. Lee , Song-Liang Chua, S.G. Johnson,
J.D. Joannopoulos and M. Soljacic, "Observation of trapped light within the
radiation continuum," Nature \textbf{499}, 188 (2013).
\bibitem{Bo Zhen}
Bo Zhen, Chia Wei Hsu, Ling Lu,  A.D. Stone, and M. Solja\v{c}i\'{c},
"Topological Nature of Optical Bound States in the Continuum,"
Phys. Rev. Lett. \textbf{113}, 257401 (2014).
\bibitem{Yang}
Yi Yang, Chao Peng, Yong Liang, Zhengbin Li, and S. Noda,
"Analytical Perspective for Bound States in the Continuum in Photonic Crystal Slabs,"
Phys. Rev. Lett. \textbf{113}, 037401 (2014).
\bibitem{Hein1}
S. Hein, W. Koch and L. Nannen, "Trapped modes and Fano resonances in
two-dimensional acoustical duct-cavity systems,"
J. Fluid Mech. \textbf{692} 257 (2012).
\bibitem{Xiong}
L. Xiong, W. Bi, and Y. Aur\'{e}gan, "Fano resonance scatterings in waveguides with impedance
boundary conditions," J. Acoust. Soc. Am. \textbf{139}, 764 (2016).
\bibitem{Hein2}
S. Hein and W. Koch, "Acoustic resonances and trapped modes in pipes and tunnels,"
J. Fluid Mech. \textbf{605}, 401 (2008).
\bibitem{Kim}
C.S. Kim, A.M. Satanin, Y.S. Joe, R.M. Cosby, "Resonant tunneling in a quantum waveguide:  Effect of
a finite-size attractive impurity,"  Phys. Rev. B \textbf{60} 10962–-10970 (1999).
\bibitem{Pilipchuk}
A.S. Pilipchuk, A.F. Sadreev, "Accidental  bound states in the continuum in an open Sinai
billiard,", Phys. Lett. A\textbf{381}, 720–724 (2017).
\end{thebibliography}
\end{document}